\documentclass[12pt,showpacs,floatfix,preprintnumbers,nofootinbib]{iopart}

\usepackage{amsopn}
\usepackage{graphicx,psfrag}
\usepackage{amssymb}
\usepackage{hyperref}
\usepackage{color}
\usepackage{graphicx}
\usepackage{acronym,xspace}
\usepackage{url}
\newcommand{\be}{\begin{equation}}
\newcommand{\ee}{\end{equation}}
\newcommand{\bea}{\begin{eqnarray}}
\newcommand{\eea}{\end{eqnarray}}
\newcommand{\bel}{\begin{align}}
\newcommand{\eel}{\end{align}}

\newcommand*{\bbar}[1]{\bar{\bar{#1}}}

\def\GMc2{{\rm G M_{\odot} c^{-2}}}

\def\PhenomPv2{\texttt{IMRPhenomPv2}\xspace}
\def\SEOBNRv3{\texttt{SEOBNRv3}\xspace}

\usepackage{color}
\definecolor{cyan}{rgb}{0,0.9,0.9}
\definecolor{orange}{rgb}{0.9,0.5,0}
\definecolor{magenta}{rgb}{1,0,1}
\definecolor{purple}{rgb}{0.8,0.4,0.8}
\definecolor{gray}{rgb}{0.5,0.5,0.5}
\definecolor{mygreen}{rgb}{0.1,0.8,0.1}
\definecolor{darkblue}{rgb}{0.0,0.0,0.6}

\newcommand{\AEIHannover}{Max Planck  Institute for Gravitational Physics
(Albert Einstein Institute), Callinstr.~38, D-30167 Hannover, Germany}
\newcommand{\UniHannover}{Leibniz Universit\"at Hannover, D-30167 Hannover, Germany}
\newcommand{\AEIPotsdam}{Max Planck  Institute for Gravitational Physics
(Albert Einstein Institute), D-14476 Potsdam-Golm, Germany}

\begin{document}

\title[Regression methods in waveform modeling: a comparative study]{Regression methods in waveform modeling: a comparative study}

%main work done by
%\author{Yoshinta \surname{Setyawati}$^{1,2}$}
%\author{Michael \surname{P\"urrer}$^{3}$}

%supporters
%\author{Frank \surname{Ohme}$^{1,2}$}

\author{Yoshinta Setyawati$^{1,2}$ \footnote{email:yoshinta.setyawati@aei.mpg.de}, Michael P\"urrer$^3$, Frank Ohme$^{1,2}$}

\address{$^1$ \AEIHannover}
\address{$^2$ \UniHannover}
\address{$^3$ \AEIPotsdam}

%\affiliation{${}^1$ \AEIHannover}
%\affiliation{${}^2$ \UniHannover}
%\affiliation{${}^3$ \AEIPotsdam}

%\date{\today}

\begin{abstract}
Gravitational-wave astronomy of compact binaries relies on theoretical models of the gravitational-wave signal that is emitted as binaries coalesce.  These models do not only need to be accurate, they also have to be fast to evaluate in order to be able to compare millions of signals in near real time with the data of gravitational-wave instruments. A variety of regression and interpolation techniques have been employed to build efficient waveform models, but no study has systematically compared the performance of these regression methods yet. Here we provide such a comparison of various techniques, including polynomial fits, radial basis functions, Gaussian process regression and artificial neural networks, specifically for the case of gravitational waveform modeling. We use all these techniques to regress analytical models of non-precessing and precessing binary black hole waveforms, and compare the accuracy as well as computational speed. We find that most regression methods are reasonably accurate, but efficiency considerations favour in many cases the most simple approach. We conclude that sophisticated regression methods are not necessarily needed in standard gravitational-wave modeling applications, although problems with higher complexity than what is tested here might be more suitable for machine-learning techniques and more sophisticated methods may have side benefits.

\end{abstract}

\maketitle

\acrodef{BH}{black hole}
\acrodef{NS}{neutron star}
\acrodef{PN}{Post-Newtonian}
\acrodef{BBH}{binary black-hole}
\acrodef{BNS}{binary neutron-star}
\acrodef{NSBH}{neutron-star black-hole}
\acrodef{EOB}{effective-one-body}
\acrodef{NR}{numerical relativity}
\acrodef{GW}{gravitational-wave}
\acrodef{PSD}{power spectral density}
\acrodef{aLIGO}{Advanced Laser interferometer Gravitational-Wave Observatory}
\acrodef{AZDHP}{aLIGO zero detuned high power density}
\acrodef{GR}{general relativity}
\acrodef{PE}{parameter estimation}
\acrodef{LAL}{LIGO algorithm library}
\acrodef{RGI}{regular grid interpolator}
\acrodef{RBF}{radial basis functions}
\acrodef{GPR}{Gaussian process regression}
\acrodef{GMVP}{greedy multivariate polynomial fits}
\acrodef{ML}{machine-learning}
\acrodef{ANN}{artificial-neural-networks}
\acrodef{MLP}{multi-layer-perceptron}
\acrodef{RBFN}{radial basis function network}
\acrodef{MCMC}{Markov-chain Monte Carlo}
\acrodef{TPI}{tensor-product-interpolation}
\acrodef{EI}{empirical interpolation}

\newcommand{\PN}[0]{\ac{PN}\xspace}
\newcommand{\BH}[0]{\ac{BH}\xspace}
\newcommand{\NS}[0]{\ac{NS}\xspace}
\newcommand{\GW}[0]{\ac{GW}\xspace}
\newcommand{\BBH}[0]{\ac{BBH}\xspace}
\newcommand{\BNS}[0]{\ac{BNS}\xspace}
\newcommand{\NSBH}[0]{\ac{NSBH}\xspace}
\newcommand{\EOB}[0]{\ac{EOB}\xspace}
\newcommand{\NR}[0]{\ac{NR}\xspace}
\newcommand{\PSD}[0]{\ac{PSD}\xspace}
\newcommand{\aLIGO}[0]{\ac{aLIGO}\xspace}
\newcommand{\AZDHP}[0]{\ac{AZDHP}\xspace}
\newcommand{\GR}[0]{\ac{GR}\xspace}
\newcommand{\PE}[0]{\ac{PE}\xspace}
\newcommand{\LAL}[0]{\ac{LAL}\xspace}
\newcommand{\RGI}[0]{\ac{RGI}\xspace}
\newcommand{\RBF}[0]{\ac{RBF}\xspace}
\newcommand{\GPR}[0]{\ac{GPR}\xspace}
\newcommand{\GMVP}[0]{\ac{GMVP}\xspace}
\newcommand{\ML}[0]{\ac{ML}\xspace}
\newcommand{\ANN}[0]{\ac{ANN}\xspace}
\newcommand{\ANNs}[0]{\acp{ANN}\xspace}
\newcommand{\MLP}[0]{\ac{MLP}\xspace}
\newcommand{\RBFN}[0]{\ac{RBF}\xspace}
\newcommand{\MCMC}[0]{\ac{MCMC}\xspace}
\newcommand{\TPI}[0]{\ac{TPI}\xspace}
\newcommand{\EI}[0]{\ac{EI}\xspace}

%===============================
%		INTRODUCTION
%===============================

\section{Introduction}
\label{sec:intro}

The laser interferometer and \GW detectors LIGO \cite{LIGO} and Virgo \cite{Virgo} have reported observations of one \BNS and ten \BBH mergers in their first two observing runs \cite{GWITC1}.
In the third observing run (O3), we expect to observe several tens of signals from compact binary coalescences \cite{dccschedule}. The analysis of these \GW data is the motivation for our study.
The data from the interferometers are filtered with many theoretically predicted waveforms with varying binary parameters. 
These waveform templates are drawn from models of the emitted \acp{GW}.
The waveform models need to fulfil accuracy and speed requirements so that the parameters of the \GW source can be estimated well in a reasonable amount of time.

We highlight two major modeling approaches: analytical and \NR.
The basis of analytical models is the \PN expansion \cite{lrrBlanchet}.
Waveform models in this category are fairly computationally efficient, but the \PN approximation breaks down for merger and ringdown part of the signal.
The second category is \NR. 
\NR waveforms are built by numerically solving Einstein's equations \cite{PhysRevLett.96.111101, PhysRevLett.95.121101, PhysRevLett.96.111102}.
Although these waveforms are known to have exceptional accuracy to model the correct \GW signals in General Relativity, they require high computational resources and need weeks to months to generate.

Combining the two approaches above, new methods have been developed to model full waveforms. 
Two major families of this group, namely the \EOB \cite{PhysRevD.64.124013, PhysRevD.78.024009, PhysRevD.95.024010, PhysRevD.95.044028} and the \emph{phenomenological} 
models \cite{PhysRevD.82.064016, PhysRevD.93.044007, PhysRevD.77.044020, PhysRevD.93.044006, PhysRevD.86.104063, PhysRevLett.113.151101} 
are commonly used in \GW analyses.
In general, these models start from a reformulation of \PN results and calibrate the model to a select number of \NR simulations.
In this study, we employ {\fontfamily{pcr}\selectfont SEOBNRv3} \cite{PhysRevD.95.024010} 
and {\fontfamily{pcr}\selectfont IMRPhenomPv2} \cite{PhysRevD.86.104063, PhysRevLett.113.151101} as two  representative models that have been widely used to explore the full parameter space of non-eccentric, precessing \acp{BBH}.

Over the past few years, complementary techniques have been developed to build fast surrogates of \EOB models and \NR waveforms with a much higher computational efficiency.
Unlike the previous approaches, these models do not start from \PN expansions.
They use existing \EOB or \NR waveforms, decompose, and interpolate them. %The results are known as the \NR surrogate family \cite{PhysRevD.96.024058, PhysRevLett.115.121102, PhysRevD.95.104036}.
The NRSurrogate models \cite{PhysRevLett.115.121102, PhysRevD.96.024058, PhysRevD.96.123011, PhysRevD.99.064045, PhysRevD.95.044028, Purrer, Purrer16} have an exceptional accuracy against the original \NR signals, but are more limited in the parameter range
and waveform length they cover.
Reduced order and surrogate models of \EOB waveforms have been crucial to allow \EOB models to be used for template bank construction \cite{PhysRevD.93.122003} and parameter estimation \cite{PhysRevD.91.042003, PhysRevD.85.104045}.

In a similar spirit, unique methods have been explored to speed up the waveform generation without compromising accuracy 
\cite{PhysRevX.4.031006, PhysRevD.95.104023, PhysRevD.99.024010, London:2018nxs, Lackey18}.
They have shown that advanced mathematical, statistical, and computational techniques are needed to build  waveform models optimized for the demands of \GW analyses.

We stress that in order to make a relatively small number of computationally expensive waveforms usable for analysis applications that rely on the ability to freely vary all parameters, all waveform models described above crucially rely on some form of 
\emph{interpolation} or fitting method as part of their construction. Phenomenological and \EOB models typically fit free coefficients (often representing unknown, higher-order \PN contributions) to a set of \NR data. %and then interpolate the values of those coefficients with some carefully chosen interpolation function. 
The fits or interpolants are then evaluated over the binary parameter space.
Other approaches, such as \NR or \EOB surrogate models, rely more on data-driven techniques to interpolate the key quantities needed to reconstruct waveforms anywhere in a given parameter-space region.
In fact, the interpolation techniques that have recently been employed cover standard methods such as polynomial fits 
\cite{PhysRevD.93.044006, PhysRevD.95.044028, PhysRevD.79.124028}, linear interpolation \cite{vinci, PhysRevD.99.024010}, 
and more complex method such as \GPR \cite{PhysRevD.96.123011, Lackey18}.
Additionally, novel interpolation methods have been developed such as \GMVP \cite{London:2018nxs, PhysRevD.95.104023} and \TPI \cite{Purrer16, Purrer}.

In this study, we investigate the importance of interpolation and fits in waveform models (which themselves are crucial for \GW astronomy), given the accuracy and computational time of various regression methods.
%in which cases should we use specific methods. 
We study whether the use of more complicated methods to model the waveforms given the same data preparation and noise reduction is justified in practice. 
Finally, we compare the performance of machine learning against various traditional methods. In particular, we explore the prospects of \ANN as a regression method \cite{paperdeep, PhysRevLett.122.211101} that has not been widely employed in waveform modeling so far. We focus
on \BBH systems
with spins either aligned with the orbital angular momentum or precessing
and provide both theoretical overviews and references to practical tools such as \emph{ready-to-use} algorithms.
Our analysis is not only of relevance for current LIGO and Virgo data and their extensions such as the Advanced LIGO A+, Voyager \cite{LIGO_white_paper}, and KAGRA \cite{kagra}, 
but also for future analysis of \GW data by LISA \cite{lisa} and the third generation instruments such as Einstein Telescope \cite{Einsteintelescope} and Cosmic Explorer \cite{Cosmicexplorer}.

The testbed we use is as follows. We compare various methods on waveform data at a fixed point in time as a function of mass ratios and spins.  
We use two models to generate waveform data: 
the time-domain model {\fontfamily{pcr}\selectfont SEOBNRv3} \cite{PhysRevD.95.024010}, and
the inverse Fourier transform of {\fontfamily{pcr}\selectfont IMRPhenomPv2} \cite{PhysRevD.86.104063, PhysRevLett.113.151101} which is natively given in the frequency domain.
Both models were designed for precessing \BBH mergers which are described by seven intrinsic parameters: the mass ratio $q$ and the two spin vectors $\vec{\chi}_{1}$ and $\vec{\chi}_{2}$ with Cartesian components in the $x, y, z$ directions.
{\fontfamily{pcr}\selectfont IMRPhenomPv2} models precessing waveforms in a single spin approximation using an effective precession spin parameter.

We consider two classes of training data:
\begin{enumerate}
\item Data on a regular three-dimensional grid describing  nonprecessing binaries, ($q, \chi_{1z}, \chi_{2z}$), where $1 \leq q \leq 10$ and $|\chi_{iz}| \leq 1$ for $i=1,2$.
\item Random uniform data on a full seven-dimensional grid ($q, \vec{\chi}_{1}$ and $\vec{\chi}_{2}$), where $1 \leq q \leq 2$ and $-1/\sqrt{3} \leq \vec{\chi}_{i} \leq 1/\sqrt{3}$ for $i=1,2$. 
\end{enumerate}
For each case, the regression methods were tested over test sets made up from random uniform test points that were drawn independently of the training set, but covering the same physical domain.

This paper is organized as follows.
We prepare the data by defining the waveform and its reference frame and defining waveform data pieces in a precession adapted frame as discussed more detail in sec.~\ref{sec:preparedata}.
We explain the background and the features of traditional methods such as linear interpolation, \TPI, polynomial fit, \GMVP, and \RBF 
as well as machine learning methods, \GPR and \ANN in section \ref{sec:interpmethods}.
In section \ref{sec:result} we present	 the results of our study.
Finally, a brief conclusion and discussion of future studies are found in section \ref{sec:discussion}.
Throughout the manuscript, we employ geometric units with the convention $G=c=1$.

%===============================
%		METHOD
%===============================

\section{Method}
\label{sec:method}

\subsection{Waveform data}
\label{sec:preparedata}

% \MP{Should use acrodef}

% \todo{
% {\color{brown}MP
% \begin{enumerate}
% \item We prepare waveform data either on 3D regular grids or scattered data in 7D for 2 waveform models
% \item Parameter ranges
% \item Explain how to prepare the data in different frames and why
% \item We focus on (how many?) key quantities
% \end{enumerate}
% }
% }

We generate training and test waveform datasets for various regression methods from two state-of-the art models of the \acp{GW} emitted by merging \acp{BBH}. We use the phenomenological model {\fontfamily{pcr}\selectfont IMRPhenomPv2}~\cite{PhysRevLett.113.151101, PhysRevD.93.044007, PhysRevD.93.044006} and the effective-one-body model {\fontfamily{pcr}\selectfont SEOBNRv3}~\cite{Pan:2013rra, Taracchini:2013rva, PhysRevD.95.024010}.
{\fontfamily{pcr}\selectfont IMRPhenomPv2} includes an effective treatment of precession effects, while {\fontfamily{pcr}\selectfont SEOBNRv3} incorporates the full two-spin precession dynamics. The models have been independently tuned in the aligned-spin sector to \NR simulations.

The \GW strain can be written as an expansion into spin-weighted spherical harmonic modes in the inertial frame
\begin{equation}
  \label{eq:strain-modes}
  h(t; \vec\lambda; \theta, \phi) = \sum_{\ell=2}^\infty \sum_{m=-\ell}^{\ell} h_\mathrm{i}^{\ell,m}(t; \vec\lambda) {}_{-2}Y_{\ell,m}(\theta, \phi).
\end{equation}
We can choose to model the waveform modes $h_\mathrm{i}^{\ell,m}(t; \theta)$ directly which depend a collection of parameters $\vec\lambda$. The spherical harmonics ${}_{-2}Y_{\ell,m}(\theta, \phi)$ for a given $(\ell,m)$ depend on the direction of emission described by the polar and azimuthal angles $\theta$ and $\phi$.
The two waveform models employed in this study provide approximations to the dominant modes at $\ell = 2$.
In a precession adapted frame {\fontfamily{pcr}\selectfont SEOBNRv3} includes $m=\pm 2$ and $m=\pm 1$ modes (the negative $m$ modes by symmetry), whereas {\fontfamily{pcr}\selectfont IMRPhenomPv2} includes only the $m=\pm 2$ modes.
For {\fontfamily{pcr}\selectfont SEOBNRv3} we directly generate time-domain inertial modes $h_\mathrm{i}^{2,m} (t)$, while for
{\fontfamily{pcr}\selectfont IMRPhenomPv2} we compute the native inertial modes in the Fourier domain $\tilde h_\mathrm{i}^{2,m} (f)$, and subsequently condition and inverse Fourier transform them to obtain an approximation to the time-domain modes.

To test interpolation methods we work in the setting of the \ac{EI} method~\cite{PhysRevD.96.024058, PhysRevX.4.031006}. In this approach we can define an \emph{empirical interpolant} of waveform data piece $X(t; \vec\lambda)$ (such as, e.g., amplitude or phase of the gravitational waveform) by
\begin{equation}
  I_N[X](t; \vec\lambda) = \sum_{i=1}^N c_i(\vec\lambda) e^i(t) = \sum_{j=1}^N X(T_j; \vec\lambda) b^j(t).
\end{equation}
The first expression is an expansion with coefficients $c_i$ of waveform data in an orthonormal linear basis $\{ e^i(t) \}_{i=1}^N$ (e.g. obtained from computing the singular value decomposition~\cite{GolubVanLoan,Demmel} for discrete data~\cite{Purrer,Purrer16}).
A transformation to the basis $\{b^i(t)\}$ allows to have coefficients which are the waveform data piece $X$ evaluated at empirical node times $T_j$. The \ac{EI} basis $\{b^i(t)\}$ and the \ac{EI} times can be obtained by solving a linear system of equations as discussed in~\cite{PhysRevX.4.031006}.
Here we forgo the basis construction step and just choose \EI times manually to select waveform data for accessing regression methods.

% # The following code should work for any waveform were we can
% # obtain time-domain inertial frame modes
%
% # Perform alignment and compute key quantities
% t, Aco22p, Aco21p, orbphase, hI_rot, qc_rot, hc_rot = \
%     align_and_compute_key_quantities(t_raw, hI_dict, t_align)
%
% Aco22p_target = spline(t, Aco22p)(t_target)
% Aco21p_target = spline(t, Aco21p)(t_target)
% orbphase_target = spline(t, orbphase)(t_target)
%
% # Save timeseries and target data
% save_key_data(t, Aco22p, Aco21p, orbphase, hI_rot, outdir+config_str)
% np.save(outdir+config_str_target,
%     [Aco22p_target, Aco21p_target, orbphase_target])

% give a brief description; we do not use the dynamics for a coarse alignment here -- that implies checking that we are not flipping between solutions

% Given an array of times `t_raw` and dictionary of
%     inertial frame modes `hI_dict` align the frame at `t_align`
%     so that \hat L_N = e_z and the phases of the (2,2) and (2,-2)
%     modes are close to zero.

% align_and_compute_key_quantities()
%   rotate_waveform
%   align_at_peak
%   align_waveform_and_frame_at_time()
%     compute_lhat
%     R_f =
%     rotate_waveform
%     rotate_waveform
%   p22 = compute_phase_at_time
%   p2m2 = compute_phase_at_time
%   pOrb = 0.25*(p22 - p2m2)
%   rotate_waveform_around_z
%   compute_phase_and_coorbital_combinations()

To simulate the process of building an efficient model we want to transform the inertial frame modes into a more appropriate form, such that data pieces are as simple and non-oscillatory as possible in time and smooth in their parameter dependence on $\vec\lambda$. 
In evaluating the model, we reconstruct the full waveforms by transforming back to the inertial frame.
This transformation includes the choice of a precession adapted frame of reference that follows the motion of the orbital plane of the binary. In this frame the waveform modes have a simple structure and are well approximated by non-precessing waveforms. A further simplification in the modes can be achieved by taking out the orbital motion. In addition, we align the waveform and frame following \cite{PhysRevD.96.024058} at the same time for different configurations and waveform models. The procedure is comprised of the following steps:\footnote{We represent rotations through unit quaternions. Quaternions can be notated as a scalar plus a vector $\mathbf{Q} = q_0 + \mathbf{q} = (q_0, q_1, q_2, q_3)$. A unit quaternion $\mathbf{R} = e^{\theta \mathbf{\hat u}/2}$ generates a rotation through the angle $\theta$ about the axis $\mathbf{\hat u}$. For calculations we use the \texttt{GWFrames}~\cite{Boyle:2013nka, GWFrames} package and notation conventions from~\cite{Boyle:2013nka}. }
% Partly from https://arxiv.org/pdf/1701.00550.pdf Sec IV
\begin{itemize}
  \item We define time relative to the peak of the sum of squares of the inertial frame modes.
  \item We transform the inertial frame waveform modes $h_\mathrm{i}^{\ell,m}(t)$ (dropping the parameter dependence on $\vec\lambda$ for now) to the minimally rotating co-precessing frame~\cite{Boyle:2011gg} and thereby obtain the co-precessing waveform modes
\begin{equation}
  h_\mathrm{copr}^{2,m}(t) = \sum_{m'} h_\mathrm{i}^{2,m}(t) \mathcal{D}^2_{m', m}\left(\mathbf{R}_\mathrm{copr}(t)\right),
\end{equation}
where $\mathcal{D}^\ell_{m', m}$ are Wigner matrices~\cite{Wigner1959, Boyle:2013nka} and $\mathbf{R}_\mathrm{copr}(t)$ is the time-dependent unit quaternion which describes the motion of this frame.
% and q(t) needs to satisfy the minimal rotation condition
% \bm{R}_\mathrm{copr}(t) has only two independent components: unit quaternion and minimal rotation condition (constrains q(t) so as to minimize the magnitude of the frame’s instantaneous angular velocity)
  \item We compute the Newtonian orbital angular momentum unit vector $\mathbf{\hat l}_N(t) = \mathbf{R}_\mathrm{copr}(t) \, \mathbf{\hat z} \, \mathbf{R}_\mathrm{copr}^*(t)$, where $\mathbf{Q}^*$ is the conjugate of the quaternion $\mathbf{Q}$ and $\mathbf{\hat z} = (0,0,1)$. We interpolate $\mathbf{\hat l}_N(t)$ to the desired alignement time $t_\mathrm{align}$.
  % align_waveform_and_frame_at_time()
  % using the SQUAD algorithm mentioned in Mike's paper
  \item We use the rotor $\mathbf{R_a} = \sqrt{- \mathbf{\hat l}_N(t_\mathrm{align}) \, \mathbf{\hat z}}$ that rotates $\mathbf{\hat z}$ into $\mathbf{\hat l_N}(t_\mathrm{align})$ to align the inertial modes at $t_\mathrm{align}$ and then compute the aligned co-precessing frame modes $\bar h_\mathrm{copr}^{2,m}(t)$ and quaternion time series $\mathbf{\bar R}_\mathrm{copr}(t)$, where the bar indicates alignment in time.
  \item Finally, we rotate around the z-axis to make the phases of the $(2,2)$ and $(2,-2)$ modes small
by applying a fixed Wigner rotation with the rotor $\mathbf{R_z} = \exp(\theta/2 \, \mathbf{\hat z})  \, \mathbf{\bar R}_\mathrm{copr}$ to obtain $\bbar{h}_\mathrm{i}^{2,m}(t)$ and $\bbar{h}_\mathrm{copr}^{2,m}(t)$.
% For simplicity we will denote the final modes obtained from this procedure without any bars.
\end{itemize}

% rotate_waveform_around_z(t, qc, hI_array, theta, t_align)
%
% p22 = compute_phase_at_time(t, hI_aligned, t_align, 4)
% p2m2 = compute_phase_at_time(t, hI_aligned, t_align, 0)
% pOrb = 0.25*(p22 - p2m2)
%% rotate_waveform_around_z(t, qc_aligned, hI_aligned, -pOrb, t_align)
%% zHat_vec = Quaternions.zHat.vec()
% R_z = Quaternions.exp(Quaternions.Quaternion(theta/2.0 * zHat_vec))
% qc_rot = R_z * qc
% _, hI_rot = rotate_waveform(t, qc_rot.conjugate(), hI_array)
% qc_rot, hc_rot = rotate_waveform(t, None, hI_rot,
%     CoprecessingToInertial=False)
% # Then take out remaining z-rotation from qc_rot

% see align_and_compute_key_quantities
% save_key_data
% t, Aco22p, Aco21p, orbphase, hI_rot

We choose the following quantities to test the accuracy and efficiency of interpolation methods:
(i) the ``orbital phase'' defined as one quarter the averaged GW-phase from the $(\ell, m) = (2,2)$ and $(2,-2)$ modes in the co-precessing frame
\begin{equation}
  \label{eq:phase}
  \phi(t) := \frac{1}{4} \left( \arg\left[\bbar{h}_\mathrm{copr}^{2,-2}(t)\right] 
                                - \arg\left[\bbar{h}_\mathrm{copr}^{2,2}(t)\right] 
                           \right),
\end{equation}
% the GW phase expected to be gauge invariant up to BMS transformations, whereas the orbital dynamics is gauge dependent
% The models we use in this study do not include mode asymmetries, so the average over m=2 and m-2 modes is not strictly necessary.
(ii) a linear combination of the $\ell = m = 2$ modes in the co-orbital frame
\begin{equation}
  %h_\pm^{2,2} = \frac{1}{2} \left( h_\mathrm{coorb}^{2,2} \pm h_\mathrm{coorb}^{2,-2*} \right),
  A(t) := \mathrm{Re} \, \bbar{h}_+^{2,2} = \frac{1}{2} \mathrm{Re} \, \left( \bbar{h}_\mathrm{coorb}^{2,2}(t) + \bbar{h}_\mathrm{coorb}^{{2,-2}^*}(t) \right),
\end{equation}
where the co-orbital modes are defined as
\begin{equation}
  h_\mathrm{coorb}^{\ell,m} (t) = h_\mathrm{copr}^{\ell,m} (t) e^{i m \phi(t)}.
\end{equation}
%h_co_22_plus = np.real(h_coorb[0] + h_coorb[4])/2
% We only use "A22" and "orbphase".

The rationale for choosing these two quantities is the following: the phasing is usually the quantity that requires the most care in \GW-modeling with accuracy requirements of a fraction of a radian over hundreds of waveform cycles. The co-orbital frame mode combinations play the role of a generalized amplitude and are typically smooth and non-oscillatory.
%\MP{I would suggest to show an example of what these key quantities look like as a function of time (for both waveforms at a single non-trivial point in parameter space) and also indicate in the figure with lines at which times we save data that for studying the performance of interpolation methods later in this paper.}

\begin{figure}[h]
\centering
     \includegraphics[width=0.46\linewidth]{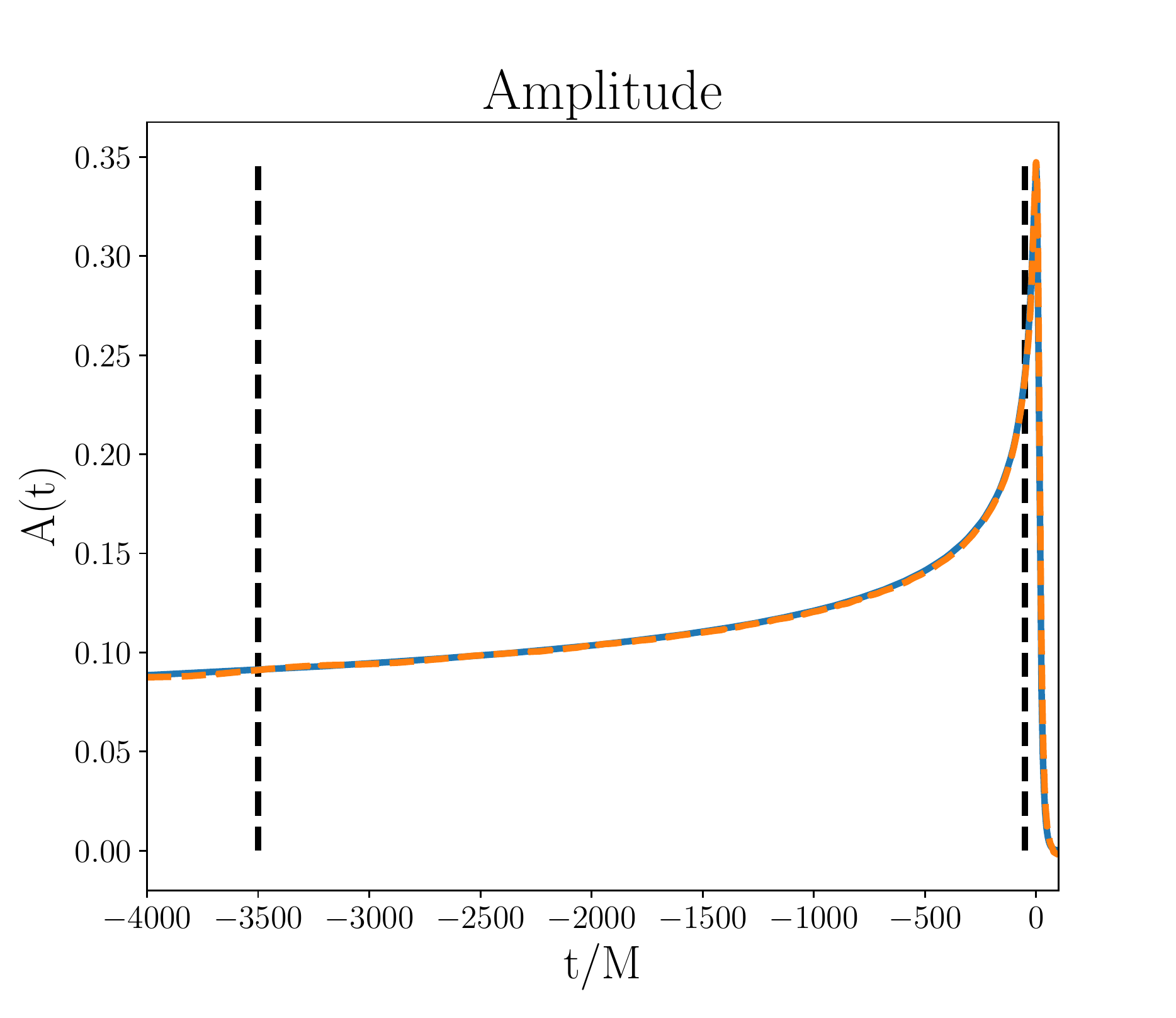}\hfil
    \includegraphics[width=0.46\linewidth]{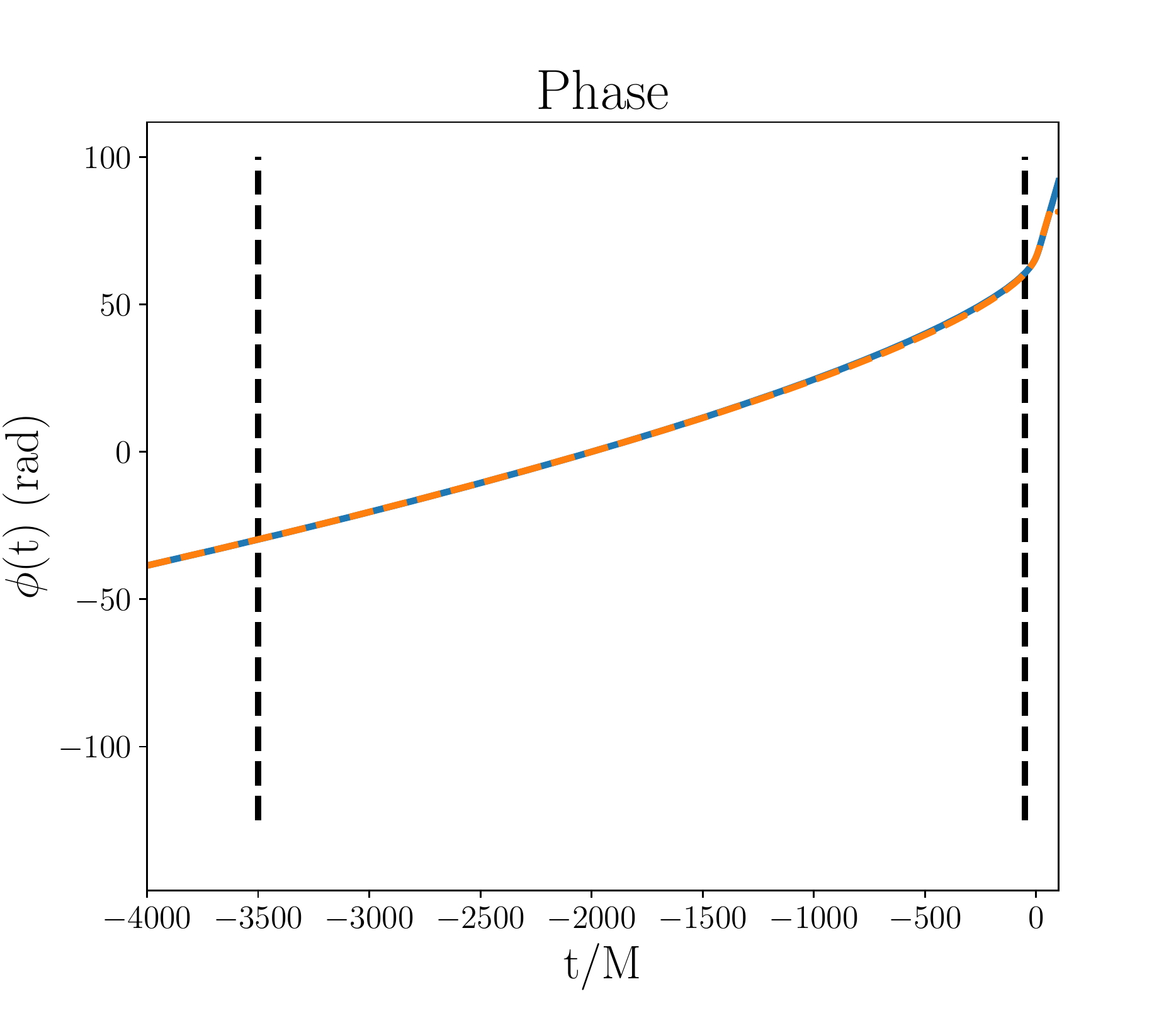}\par\medskip
    \includegraphics[width=0.5\linewidth]{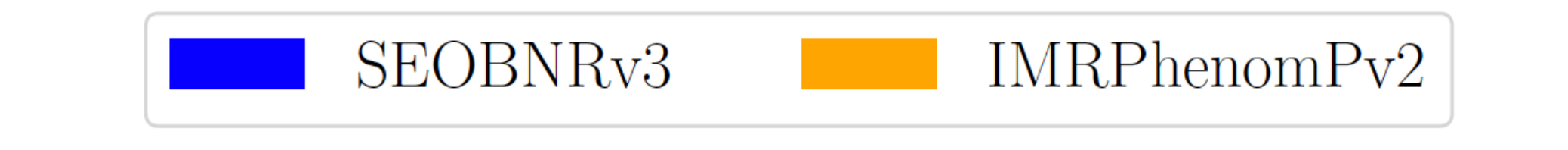}
\caption{The key quantities of the \ac{GW} signal of a precessing \BBH, here illustrated for a binary with $(q, \chi_{1x}, \chi_{1y}, \chi_{1z}, \chi_{2x}, \chi_{2y}, \chi_{2z})=(1.99, 0.51, 0.04, 0.03, 0.01, 0.6, 0.1)$.
Left: the dimensionless amplitude $A(t)$. 
Right: the phase $\phi(t)$ (in unit radian). 
The black dashed lines show the points in time-space, where we perform different interpolation methods (t=-3500M and t=-50M).}
\label{fig:method_amp_phase}
\end{figure}

We consider the following waveform training datasets in this study: (i) Three-dimensional datasets: Several interpolation methods we consider in this study require data on a regular grid. We prepare three-dimensional datasets $(q, \chi_{1z}, \chi_{2z})$ in the mass-ratio $q = m_1 / m_2$ and the aligned component spins $\chi_{iz} = \vec S_i \cdot \hat L_N / m_i^2$ for $i=1,2$. We do not include the total mass since it can be factored out from the waveform for \acp{GW} emitted from \acp{BBH} which are solutions of Einstein's equations in vacuum. The grids have an equal number of points per dimension, ranging from 5 to 11. We choose parameter ranges $1 \leq q \leq 10$ and $|\chi_{iz}| \leq 1$.
(ii) The full intrinsic parameter space we consider is seven-dimensional: we include the dimensionless spin vector of each black hole $\chi_{i} = S_i / m_i^2$ and the mass-ratio $q$ of the binary. Due to the \emph{curse of dimensionality} regular grid methods require a prohibitive amount of data in 7D. For instance, ten points per dimension would require $10^7$ waveform evaluations. Therefore, we only produce scattered waveform data in seven dimensions which are drawn from a random uniform distribution in each parameter.
Here we choose parameter ranges $1 \leq q \leq 2$ and $-1/\sqrt{3} \leq \vec\chi_i \leq 1/\sqrt{3}$.
For both choices of dimensionality we also generate test data of 2500 points drawn randomly from the respective parameter space.

Waveform data in three and seven dimensions is produced at a total mass of $M = 50 M_\odot$ with a starting frequency of $20 \mathrm{Hz}$. We align the waveform and frames at $t_\mathrm{align} = -2000 M$ with the above procedure. We record waveform data from the key quantities at two different times, $t_\mathrm{target} = -3500 M$ and $-50 M$, where we have performed alignment in time such that the mode sum of the waveform amplitudes peaks at $t = 0 M$. This choice allows us to independently probe the inspiral and the merger regime. We expect that the waveform data will be very smooth in the inspiral, but more irregular close to merger due to the calibration of internal model parameters to numerical relativity waveforms at a limited number of points in parameter space.

\subsection{Regression methods: a general overview}
\label{sec:interpmethods}

A large number of techniques have been developed to improve the speed and accuracy of generating gravitational waveforms.
A priori, one would expect that higher speed would go hand-in-hand with less accuracy and less complexity.
One frequent question is how to select a method for a specific purpose.
Depending on the goals, a choice needs to be made between complex, highly accurate methods with moderate efficiency versus simpler but more efficient methods, and we can choose to trade accuracy for speed.

In this subsection, we discuss various methods and categorize them into two groups. 
The first group is comprised of traditional interpolation and fitting methods which are based on mathematical techniques and algorithms 
that are straightforward to implement and easily evaluated.
The second group is made up of \ML methods which may require a more advanced mathematical and computational background.
Methods from the second group are in general more complex and require more computational resources than the first group.
Here we give a basic description of these methods, their limitation and provide some references.

\subsubsection{\textbf{Traditional interpolation and fitting methods}}
\label{sec:traditional}
%The objective of the traditional methods is to understand the logic and mathematical background from the input to the output of the training data.
The traditional interpolation and fitting methods are either interpolatory, i.e., the approximation is designed such that it exactly includes the data points, or they produce an approximate fit, where a distance function between
the data and the model is minimized. Many of these methods rely on polynomials as building blocks to model the data.
Some models have a fixed order of approximation, while others let the number of terms be a free parameter.
These methods are relatively straightforward to use and do not usually require much computational power.
%In most cases, the methods that fall into this group are relatively more straightforward than the \ML group. 
%Typically, these methods do not require a higher-level computational skill, making it practical and easy to use.

\begin{enumerate}
\item \textbf{Linear interpolation}
\label{sec:linear}

Linear interpolation is a straight line approximation that predicts the value of an unknown data point which lies between two known points \cite{garrido}.
Given its simplicity, this method has been widely used as a standard method to perform interpolation in various fields.
If we have several data points, the transition between the adjacent data points is only continuous but not smooth.
Higher order methods such as cubic interpolation 
can be used if a smoother approximation is desired (see subsection \ref{sec:tpi}).

Since linear interpolation is available as a standard Python package, 
we include this method to compare to other more complicated techniques. 
In particular, we investigate the application of multivariate linear interpolation on a regular grid using the \ac{RGI} \cite{rgigit, rgi} that is available in {\fontfamily{qcr}\selectfont scipy}. 

The mathematical background of linear interpolation can be explained as follows.
Assume two known points $(x_0, y_0)$ and $(x_1, y_1)$ and an unknown point $(x, y)$ with $x_0 \leq x \leq x_1$.
This method assumes that the slope between $x_0$ and $x$ is equal to the slope between $x$ and $x_1$.
Hence, we use the following relation to predict the data point $y$ in one dimension. %\footnote{Interpolation in one dimension is normally called univariate interpolation.}:
\be
\label{linearinterpolation}
\eqalign{\frac{y-y_0}{x-x_0} =\frac{y_1-y}{x_1-x} \cr
\Leftrightarrow y =y_0+(x-x_0)\frac{y_1-y_0}{x_1-x_0}.}	
\ee

In dimensions $d>1$, this method requires a regular grid of data points as a training set.
%(A refined illustration of multivariate linear interpolation in three dimensions can be found in Fig 2 of \cite{linearinter}).

Multivariate linear interpolation works as follows. 
Let $y_i (\vec{x})$ be the data point we want to predict, where $\vec{x}$ denotes the input parameters in $d$ dimensions.
Initially, we need to obtain the parameters of the projection of $y_i (\vec{x})$ in $d-1$ dimensions, 
followed iteratively by $d-2$ and so on until we reach one-dimensional case $d=1$.
Once we obtain these projection points, we can employ Eq (\ref{linearinterpolation}) to predict the values of these points in one dimension.
Subsequently, we use the predicted values as the known points to predict the result in higher dimensions iteratively.
We then repeat the process further to find $y_i (\vec{x})$ in $d$ dimensions.
This algorithm involves a small number of multiplications and additions, which are relatively fast. 

Since \RGI assumes a regular grid, it is affected by \emph{the curse of dimensionality}: the number of training points grows as the power of $d$.
Therefore, we only investigate this method in three dimensions.

Other popular regression methods that we do not consider here are ridge regression \cite{ridgeregression}, LASSO regression \cite{lassoregression}, and Bayesian regression \cite{wakefield2013bayesian}.
One reason is that the \GW training data is quite well-behaved and does not usually include outliers such that would require special treatment.

\item  \textbf{Tensor product interpolation}
\label{sec:tpi}

On regular or Cartesian product grids one can use the same univariate interpolation method in each dimension and the grid points can be unequally spaced.
This gives rise to \TPI methods. Popular choices for the univariate method are splines~\cite{deBoor} and, if the data is very smooth, spectral interpolation~\cite{Boyd2000, CanutoVolI}.

Let us assume that we want to model a waveform quantity $X(t; \vec\lambda)$ at a particular time $t=t_\mathrm{i}$. We define the $d$-dimensional \TPI interpolant (where $d = \dim(\vec\lambda)$) as an expansion in a tensor product of one-dimensional basis functions $\Psi_j(\lambda_j)$,
\begin{equation}
  % I[X](t_\mathrm{i}; \vec\lambda) = \sum_{j_1, \dots, j_n} a_{j_1, \dots, j_n}
  % \left( \Psi_1 \otimes \dots \otimes \Psi_n \right)_{j_1, \dots, j_n}(\vec\lambda),
  I[X](t_\mathrm{i}; \vec\lambda) = \sum_{j_1, \dots, j_d} a_{j_1, \dots, j_d}
  \left( \Psi_{j_1} \otimes \dots \otimes \Psi_{j_d} \right)(\vec\lambda),
\end{equation}

A popular choice for the basis functions are univariate splines, which are piecewise polynomials of degree $k - 1$ (order $k$) with continuity conditions. For instance, cubic splines have degree $k=4$ and continuous first and second derivatives. The boundaries of the domain require special attention. A simple choice is the natural spline where the second derivative is set to zero at the endpoints. If boundary derivatives are not known it is better to use the so-called ``not-a-knot'' boundary condition~\cite{deBoor}. This condition is defined by demanding that even the third derivative must be continuous at the first and last knots.

To construct splines in a general manner it is advantageous to introduce basis functions with compact support, so-called \emph{B-splines}. We denote the $i$-th B-spline basis function~\cite{deBoor, Quarteroni2000} of order $k$ with the knots vector 
%$\vec t = (t_j)_{j=1}^{n+k+1}$, with $n \in \mathbb{N}$,
$\vec t$, a nondecreasing sequence of real numbers, evaluated at $x$ by $B_{i,k,t}(x)$. The knots refer to the locations in the independent variable where the polynomial pieces of B-spline basis function are connected. For distinct knots $t_i, \dots, t_{i+k+1}$, the B-splines can be defined as
\begin{equation}
  B_{i,k,t}(x) := (t_{i+k} - t_i) [t_i, \dots, t_{i+k}] ( \cdot - x)_+^{k-1},
  % \quad \forall x \in \mathbb{R},
\end{equation}
where $[t_i, \dots, t_{i+k}] f$ is the \emph{divided difference}~\cite{deBoor, Quarteroni2000} of order $k$ of the function $f$ at the sites $t_i, \dots, t_{i+k}$, and $(x)_+:=\max\{x,0\}$. The B-splines can also be defined in terms of recurrence relations. The definition can be extended to partially coincident knots which are useful for the specification of boundary conditions.
B-splines can be shown to form a basis~\cite{deBoor} of the spline space for a given order and knots vector.
A spline function or spline of degree $k$ with knots $\vec t$ can be then defined as an expansion
% I avoid saying what the dimensionality of the spline space $\bm{\mathcal{S}}_{k,\Delta}$ is and to define this space in the first place since it gets a bit too involved for this paper. Assuming a particular grid $\Delta = t_0 < t_1 < \dots t_{l+1}$ one can define an extended grid with the boundary nodes being conincident $k$ times. Then $n = l +k = \dim \bm{\mathcal{S}}_{k,\Delta}$ (see e.g. Deuflhard & Hohmann, p. 223). Then $f = \sum_{i=1}^n c_i B_{i,k,t}$.
\begin{equation}
  s = \sum_{i} s_i B_{i,k,t}(x),
\end{equation}
with real coefficients $\{s_i\}_{i=1}^n$. Given data, a fixed order and knots vector, and a choice of boundary conditions, we can solve the linear system for the spline coefficients $s_i$. For efficient evaluation we only compute the parts of the B-spline basis functions that are nonzero.
% Would be good to say what the computational complexity is at least for evaluation.

% The above-mentioned ``not-a-knot'' boundary condition can be imposed as follows at the first knot $x_1$
% \begin{equation}
% B_{i,k,t}^{(3)} \left(\frac{x_0 + x_1}{2} \right) \overset{!}{=} B_{i,k,t}^{(3)} \left(\frac{x_1 + x_2}{2}\right)
% \end{equation}

For smooth data, Chebyshev interpolation~\cite{Boyd2000, CanutoVolI} is a popular choice. Chebyshev polynomials (of the first kind) are defined as the unique polynomials satisfying 
\begin{equation}
  T_n(\cos(\theta)) = \cos (n \theta)
\end{equation}
on [-1,1].
In contrast to splines where the polynomial degree is usually low, global high order polynomial interpolation requires a special choice of nodes to be well-conditioned. A good choice are Chebyshev-Gauss-Lobatto nodes (which are defined to be the extrema of the $T_n(x)$ plus the endpoints of the domain)
\begin{equation}
  x_k = -\cos \left( \frac{k\pi}{m-1} \right), \quad k = 0,\dots, m-1.
\end{equation}
Then we can approximate a function $f(x)$ by an expansion
\begin{equation}
  f(x) \approx I[f(x)] := \sum_{k=0}^{m-1} c_k T_k(x),
\end{equation}
For $f \in C^\infty$ the error of Chebyshev interpolation converges exponentially with the number of polynomials $T_n(x)$.

Tensor product interpolation is a very useful tool for constructing fast \emph{reduced order models} (ROM) or \emph{surrogate models} of time or frequency dependent functions that depend on a moderate number of parameters $\vec\lambda$. \TPI with splines and Chebyshev polynomials has been used to build several \GW models~\cite{PhysRevD.96.123011, PhysRevD.95.104023, Purrer16, Purrer} and~\cite{Lackey:2016krb}, respectively.
\TPI is not available in standard Python packages. For \TPI spline interpolation we use the Cython~\cite{Cython} implementation in the {\fontfamily{qcr}\selectfont TPI} package~\cite{TPI}.

\item \textbf{Polynomial fits}
\label{sec:polyfit}

A polynomial fit is a multiple linear regression model where the independent variables form a polynomial \cite{Polynomialbook}.
%The degrees and constants of the polynomials are set to fit the data.
Different settings of maximum polynomial degrees may cause \emph{underfitting} or \emph{overfitting}, therefore care must be taken in choosing the ansatz.

%The description of the polynomial fit is as follows.
Assume that we have $N$ training points $(\{\vec{x}_i, y_i\}  \in \mathbb{R}^d \times \mathbb{R}|i=1, \cdots, N)$.
Our goal is to find a function or regressor such that each $\vec{x}_i$ yields an output with the lowest error against its function values $y_i$. 
We assume that this function $f(\vec x)$ is expressed by a polynomial of degree $k$ and parameters $\vec{c}$. 

In one dimension we have:
%Let $f(\vec{x})$ be the function we want to predict (the interpolant), where $\vec{x}$ depends on a set of parameters $d$.
%We want to find the interpolant given $N$ training points, where the output of each data point ($\vec{y_i}$) is a function of $f(\vec{x_i})$.
%We assume that $f(\vec{x})$ is expressed by a set of polynomial degree and constants as follows:
\be
\label{polyfunc}
f(\vec x)=c_0 x^k + c_1 x^{k-1} + \cdots + c_{k-1} x + c_k .
\ee
If we had as many degree of freedom as data points, we could demand:
\begin{equation}
\label{polyfxy}
f(x_i)=y_i.
\end{equation}
In matrix form, Eq (\ref{polyfxy}) can be written as:
\begin{eqnarray}
\label{polyfit}
\fl 	 &\mathbf{X}\vec{c} = \vec{Y} \nonumber\\
	 &\left({\begin{array}{ccccc} 
	x_1^k & x_1^{k-1}  & \cdots & x_1 & 1 \\
	x_2^k & x_2^{k-1}  & \cdots & x_2 & 1 \\
	\vdots &  & \ddots &  & \vdots \\
	x_N^k & x_N^{k-1}  & \cdots & x_N & 1 \\
	\end{array}}\right)
	\left(\begin{array}{c} c_0 \\ c_1 \\ \vdots \\ c_k \end{array}\right)	
	=
	\left(\begin{array}{c} y_1 \\ y_2 \\ \vdots \\ y_N \end{array}\right),
\end{eqnarray}
where \textbf{X} is the $N \times (k+1)$ Vandermonde matrix.
The parameters $\vec{c}$ are obtained by solving Eq (\ref{polyfit}) for the known input and output data, \textbf{X} and $\vec{Y}$ in the training set.
In general, the linear system may be over or under determined such that no unique solution would exist. 
Instead, we employ the standard discrete least squares fit to minimize the error (see section 10 of \cite{polybook} and \cite{numC}):
\be
\Sigma_{j=1}^N |{f(x_j)-y_j}|^2
\ee
Similar to linear interpolation, univariate polynomial interpolation is available in the {\fontfamily{qcr}\selectfont scipy} package.

%Multivariate interpolation deals with interpolation of a function of more than one parameter ($d>1$).
Ref \cite{Polynomialbook} discusses several methods and provide an overview of multivariate interpolation with polynomials.
We employ polynomial fits for multivariate interpolation as in \cite{rompy} and explained more detail in \cite{Field:2011mf}.

%Here we employ the same procedure as described above for multivariate interpolation,
%except that we modify the Vandermonde matrix $\mathbf{X}$ such that it does not only 
%depend on a number of polynomial degree $D$, but also to a number of parameter $d$.

%We express the $N \times Z$ Vandermonde matrix $X_{ij}$ as:
%\be
%\label{modVan}
%	 \left({\begin{array}{ccccc} 
%	1 & \phi_{11} & \phi_{12} & \cdots & \phi_{1z} \\
%	1 & \phi_{21} & \phi_{22} & \cdots & \phi_{2z} \\
%	\vdots &  & \ddots &  & \vdots \\
%	1 & \phi_{n1} & \phi_{n2} & ... & \phi_{nz} \\
%	\end{array}}\right)
%	\left(\begin{array}{c} c_0 \\ {c_1} \\ {c_2} \\ \vdots \\ c_{D-1} \end{array}\right)	
%	=
%	\left(\begin{array}{c} y_n \\ y_{n-1} \\ \vdots \\ y_1 \end{array}\right),
%\ee

%For ($d>1$), we employ {\fontfamily{qcr}\selectfont polyfitnd} in \cite{rompy}.

\item \textbf{Greedy multivariate polynomial fit (GMVP)}

London and Fauchon-Jones \cite{London:2018nxs} recently introduced methods
that build an interpolant for a given data set by adaptively choosing a small
set of analytical basis function from a certain class of functions. In our
study here, we test the \GMVP procedure described in detail in Sec.~II B of
\cite{London:2018nxs}.

In this method, a scalar function, $f$, that is known at discrete points in
the $d$-dimensional parameter space, $\vec x_j = \{ x^1_j,
x^2_j, \ldots, x^d_j \}$, is approximated by a linear sum of
analytical basis functions, $\phi_k(\vec x)$,
\be
 f(\vec x) \approx \sum_k \mu_k \, \phi_k(\vec x).
\label{eq:gmvp_basis}
\ee
Given a set of basis functions, the coefficients $\mu_k$ are determined by a
`least-squares' optimal fit to the known function values $f(\vec
x_j)$. In practice, this is calculated using the pseudoinverse
(Moore-Penrose) matrix of $\phi_k(\vec x_j)$ (that is, the values of the
basis functions at the given location in the parameter space).

In \GMVP, the basis functions are chosen to be multivariate polynomials of
maximal degree $D$. In order to prevent overfitting, however, not all possible
polynomial terms from the set
\be
 \phi_k(\vec x) \in \left \{ \big(x^1 \big)^{\alpha_1}
\,\big(x^2 \big)^{\alpha_2}  \ldots \big(x^n\big)^{\alpha_d}, \quad
\sum_{i=1}^n \alpha_i \leq D \right\}
\ee
are included in the basis. Instead, a greedy algorithm \cite{Field:2011mf}
iteratively adds the basis functions to (\ref{eq:gmvp_basis}) that minimize the
error
\be
 \epsilon^2 = \frac{\sum_j \left[ f(\vec x_j) - \sum_k \mu_k \,
\phi_k(\vec x_j) \right]^2}{\sum_j \left[ f(\vec x_j) \right]^2} .
\ee
This process terminates when the difference in $\epsilon$ between two
successive iterations becomes smaller than some user-defined
tolerance. In order to improve the stability of the algorithm, the maximally
allowed multinomial degree $D$ is successively increased, which the authors
of \cite{London:2018nxs} refer to as degree tempering.

In our study, we use \GMVP with a tolerance of $\epsilon = 5 \times 10^{-4}$
and a maximal multinomial degree of $D = 16$.

\item \textbf{Radial basis functions (RBF)}
\label{sec:rbf}

Radial basis functions \cite{Buhmann} are an approximation for continuous functions, where the predicted outputs depend on the Euclidean distance between the points and a chosen origin.
This method is applicable in arbitrary dimensions and does not require a regular grid. 

We include \RBF in this study due to several reasons.
Primarily, because this method is simple, rapid, and has been integrated as a standard Python package in {\fontfamily{qcr}\selectfont scipy}.
Moreover, \acp{RBF} are used in machine learning as activation functions in radial basis functions neural networks (see section \ref{sec:ml}).  
%\RBF (also known as the \emph{single-layer perceptron}) is the simplest form of the artificial neural networks (ANN).

The mathematical background of \acp{RBF} is explained as follows.
Let $N$ be the number of training points, $\vec{x}_i$ the parameters of each data point, and $y_i$ the data defining the training set $\{(\vec{x}_i, y_i) \in \mathbb{R}^d \times \mathbb{R}|i=1, \ldots, N\}$.

The goal is to find an approximant $s:\mathbb{R}^d \to \mathbb{R}$ to the function $y:\mathbb{R}^d \to \mathbb{R}$ 
such that $s(\vec{x}_i)=y_i$ ($s$ interpolates $y$ at the chosen points) with the form: 
%For consistency with the notation in the previous section, $\vec{x}$ is a vector of parameters.
\be
	s(\vec{x}) = \sum_{i=1}^N w_i \varphi (r), 
\ee
where $\vec{x}$ is the vector of independent variables, $w_i$ are the coefficients, $r$ is the Euclidean distance between $\vec{x}$ and $\vec{x}_i$ ($r=\|\vec{x}-\vec{x_i}\|$),
 and $\varphi(r)$ is known as the \emph{radial basis function}.

To obtain the approximant $s$, we need to solve: 
\be
	\Phi (r)\vec{w}=\vec{Y},
\ee
where $\Phi (r)=\{\|\vec{x}-\vec{x_i}\|\}_{x, x_i \in \Xi}$, $\vec{Y}=\{y_i\}_{i=1}^N$ 
and $\vec{w}=\{w_i\}_{i=1}^N$. 
$\Xi$ is a finite subset of $\mathbb{R}^d$ with more than one element \cite{Buhmann}.
We can solve the linear system for the coefficients and obtain the interpolant. 
Hence, the computational complexity and thus the training time of \RBF is dominated by the computation of vector coefficients $\vec{w}$
that involves matrix inversion and goes as $\mathcal{O} (N^3)$ \cite{Rbf5}.

%\begin{eqnarray}
%\label{rbf}

%\begin{split}
%\fl	 \left(\begin{array}{c} y_1 \\ y_2 \\ \vdots \\ y_n \end{array}\right) = 
%	\left(\begin{array}{c} w_{1} \\ w_{2} \\ \vdots \\ w_n \end{array}\right)\\
%	 \left({\begin{array}{cccc} 
%	\varphi(\|\vec{x_1}-\vec{x_1}\|) & \varphi(\|\vec{x_1}-\vec{x_2}\|) & \cdots & \varphi(\|\vec{x_1}-\vec{x_n}\|) \\
%	\varphi(\|\vec{x_2}-\vec{x_2}\|) & \varphi(\|\vec{x_2}-\vec{x_2}\|) & \cdots & \varphi(\|\vec{x_2}-\vec{x_n}\|) \\
%	\vdots &  & \ddots & \vdots  \\
%	\varphi(\|\vec{x_n}-\vec{x_1}\|) & \varphi(\|\vec{x_n}-\vec{x_2}\|) & \cdots & \varphi(\|\vec{x_n}-\vec{x_n}\|) \\
%	\end{array}}\right).
%\end{split}
%\end{eqnarray}

The interpolation matrix $\Phi (r)$ has to be nonsingular so that it does not violate the Mairhuber-Curtis theorem \cite{Buhmann}.
The solution is to choose a kernel function such that $\Phi(r)$ is a semi-definite matrix and therefore nonsingular.
One common choice is the multiquadric kernel function $\varphi(r)$ expressed by:
\begin{equation}
\varphi(r)=\sqrt{1+\bigg(\frac{r}{\varepsilon}\bigg)^2},
\end{equation}
where $\varepsilon$ is the average distance between nodes based on a bounding hypercube as defined in {\fontfamily{qcr}\selectfont scipy} \cite{rbf}. 

The multiquadric kernel function is commonly applied to scattered data because of its versatility due to its adjustable parameter $\varepsilon$ which can improve the accuracy or the stability of the approximation.
Ref.~\cite{Buhmann} shows that this kernel is also able to approximate smooth functions well so that it useful for approximation. 
Hence, we employ the \emph{multiquadric} kernel function in this study.
\end{enumerate}

\subsubsection{\textbf{Machine learning methods (ML)}}
\label{sec:ml}

Machine learning is the scientific study of computer algorithms and statistics which aims to find patterns or regularities in the data sets.
Systems learn from the training data and can predict output values for test data.
\ML is a branch of artificial intelligence.

Although the distinction is a blur, one major difference between \ML and traditional interpolation methods lies in their objectives.
In traditional methods, the objective is not only to provide an approximation of an underlying function from which the training data were generated, 
but also to understand the mathematical process behind the relation of input and output data.
In that case, we seek interpolants or fits which often can be found analytically by solving linear systems for the coefficients in the model.
Hence, the traditional methods originated from approximation theory and numerical analysis in mathematics.
Conversely, in machine learning, the objective is to recognize patterns from the input-output training set and to construct a model from this data.
Although we know that the result follows some mathematical procedures that depend on free parameters, 
these details are considered to be less important.
Hence \ML can be seen as a sub-field of computer science.

\begin{enumerate}
\item \textbf{Gaussian process regression (GPR)}

\GPR is a unique method that combines statistical techniques and machine learning.
It can predict function values away from training points and can provide uncertainties of the predicted values,
which will be useful for certain applications.
\GPR can be used with multivariate scattered data.

Compared to traditional methods, \GPR requires more knowledge of advanced statistics such as covariance matrices, regression and Bayesian statistics for the optimization strategy.
\GPR can be considered as a combination of traditional and machine learning methods. 
%and the parametric and nonparametric regression model that will be discussed later in this subsection.

We provide a summary of \GPR as discussed in detail in Ref.~\cite{gpr06, scikit-learn}.
%In this study, we follow the notation and theoretical background of \GPR in our main references \cite{gpr06, scikit-learn}, 
%where they provide a detailed study and a more explanation for further applications.
%We categorize \GPR in \ML group because some parts of the methods used in \GPR have shown to have \ML characteristics.
%\GPR is a nonparametric regression model, meaning that it does not assume that the data distribution is defined by a finite set of parameters. 
%As an approach to perform interpolation, \GPR has more complexity than the previous methods.
%Various groups have employed \GPR to investigate its prospect to build a waveform model \cite{PhysRevD.96.123011}.
We start with the most important assumption in \GPR. Any discrete set of function values $y_i =  y(\vec{x_i})$ is assumed to be a realization of a Gaussian process (GP).
Assuming the data can be pre-processed to have zero mean, $\mu(\vec{x})=0$, the covariance function $k(\vec{x}, \vec{x}')$ fully defines the Gaussian process:
\begin{equation}
y(\vec{x}) \sim GP\bigg(\mu(\vec{x})=0, k(\vec{x}, \vec{x}')\bigg).
\end{equation}

%The covariance function $k(\vec{x}, \vec{x}')$ is defined by some kernel functions and \emph{hyperparameters} $\vec{\theta}$. 
%The kernel function should not be randomly selected as some features will affect the values of the result.
%We will discuss some features and differences of \GPR kernels as well as the optimization of the hyperparameters $\vec{\theta}$ later in the text.
Assume that we want to predict the value $y_*$ at $\vec{x}_* \in \mathbb{R}^d$
and that we have $N$ numbers of training points, where each point depends on $d$ parameters expressed by $\{(\vec{x}_i, y_i)|i=1, \ldots, N\}$. 
%The values of these points are expressed by $\vec{y} \in \mathbb{R}^{N \times 1}$. 
The training and test outputs can be written as follows:
\begin{equation}
\left[{\begin{array}{c} 
	\vec{y} \\
	y_* 
\end{array}}\right]
\sim
\mathcal{N} 
\left( 0, \left[{\begin{array}{cc} 
	K (X, X)+\sigma_n^2 I & K (X, X_*) \\
	K (X_*, X) & K (X_*, X_*) 
\end{array}}\right]
\right),	
\end{equation}
where %$X$ denotes $N$ training data of $\vec x \in \mathbb{R}^d$, and similarly $X_*$ for $N_*$ of test data $\vec x_* \in \mathbb{R}^d$,
$K (X, X)$ denotes the matrix of the covariances evaluated at all pairs of the training points and similarly for $K(X_*, X_*)$, $K(X, X_*)$, and $K(X_*, X)$,
$\sigma_n^2$ (also called \emph{nugget}) is the variance of the Gaussian (white) \emph{noise} kernel that will be discussed later (see the \emph{hyperparameters}).

Explicitly, in order to predict a single value $y_*$, we need to compute $K(X, X)$ as the covariance matrix between each point in the training set,
$K (X, X_*)$ and its transpose that are vectors and the scalar $K (X_*, X_*)$.
In a different form, our main goal is to find the conditional probability expressed by the following distribution:
\begin{equation}
p(y_* | \vec{x}_i, \vec{x}_*, \vec{y}, \vec{\theta}) = \mathcal{N} (\bar{y}_*, \textrm{var} (y_*)),
\end{equation}
i.e., the probability of finding the value $y_*$ given the training data $\vec{x}_i$ and $ \vec{y}$,
the hyperparameters $\vec{\theta}$, and the location $\vec{x}_*$  
is a normal distribution with mean $\bar{y}_*$ and variance $\textrm{var} (y_*)$.

The mean and variance can be shown to be:
\be
\label{meanvar}
\bar{y}_* =  K ( X_*, X) (K (X, X))^{-1}_{ij} y_j %\numberthis \\
\ee
\be
\mathrm{var} (y_*) = K (X_*,  X_*) - 
K (X_*, X_i) (K(X, X))^{-1}_{ij} K (X_*, X_j). %\numberthis
\ee
In the equation above, the covariance $K(x_i, x_j)$ is expressed by:
\be
\label{eq:cov}
K(x_i, x_j)=\sigma_f^2 k(x_i,x_j)+\sigma_n^2 \delta_{ij},
\ee
where 
$\sigma_f$ and $\sigma_n$ are hyperparameters, 
$\delta_{ij}$ is the standard Kronecker delta, 
$k(x_i, x_j)=k(r)$, %where $k$ denotes the covariance function (see later in the text), 
and $r$ is the distance:
\be
\label{eq:gpr_r}
r=\sqrt{(\vec{x}-\vec{x}')^T M(\vec{x}-\vec{x}')}.
\ee
In the following, we discuss the form of $M$ as a diagonal matrix with a tunable length scale in each physical parameter
which form part of the hyperparameters.
\\

\textbf{The hyperparameters}

We assume that our training data has some numerical noise $\sigma_n^2$ and a scale factor $\sigma_f$ 
that can be estimated by optimizing the hyperparameters $\vec{\theta}=\{\sigma_f, \sigma_n, M\}$.
For instance, the explicit form of $M$ in the seven-dimensional case is:
\be
M=\textrm{diag}(\ell_q^{-2}, \ell_{\chi_{1x}}^{-2}, \ell_{\chi_{1y}}^{-2}, \ell_{\chi_{1z}}^{-2}, \ell_{\chi_{2x}}^{-2}, \ell_{\chi_{2y}}^{-2}, \ell_{\chi_{2z}}^{-2}),
\ee
where the $\ell_i$ are length scales. Ref.~\cite{gpr96} describes the length-scale $\ell$ as the distance taken in the input space before the function value changes significantly.
Small values of the lengthscale $\ell$ imply that the function values change quickly and vice versa.
%Furthermore, it determines how we can extrapolate the training data.
Hence, the lengthscale $\ell$ describes the smoothness of a function.

To determine the hyperparameters, we can maximizse the marginal log-likelihood:
\begin{eqnarray}
\label{eq:gprlog}
\fl \ln p(y_i|\vec{x}_i,\vec{\theta}) = -\frac{1}{2}\bigg(y_i(K(X, X))^{-1}_{ij}y_j 
 + \ln |K(X, X)| + N \ln 2\pi \bigg). 
\end{eqnarray}
%where $N$ is the number of training points following our notation.

Because the log-likelihood may have more than one local optimum, 
we repeatedly start the optimizer and we choose ten repetitions.
For the first run, we set the initial
value of each length scale to unity, with bounds of $10^{-5}$ to $10^5$. 
Furthermore, we set $\sigma_n^2=10^{-10}$, where higher $\sigma_n^2$ value means that the data are more irregular.
The subsequent runs use the allowed values of the hyperparameters from the previous runs until the maximum number of iterations is achieved.

In Eq (\ref{eq:gprlog}), we see that the partial derivatives of the maximum log likelihood can be computed using matrix multiplication. 
However, the time needed for this computation grows with more data in the training set
as $\mathcal{O } (N^3)$. 
Additionally, we employ Algorithm 2.1 of \cite{gpr06}, because Cholesky decomposition
is about six time faster than the ordinary matrix inversion to compute Eq (\ref{eq:gprlog}).
We highlight that although \GPR becomes more accurate in predicting the underlying functional form of the data given more training points $N$, 
it has complexity $\mathcal{O}(N^3)$ and therefore the method becomes ineffective for large $N$.

We estimate the posterior distribution of the hyperparameters using Bayes' theorem as follows:
\be
p(\vec{\theta}|\vec{x}_i,y_i) \propto p(\theta) p(y_i|\vec{x}_i,\vec{\theta}),
\ee
where we employ a uniform prior distribution $p(\theta)$.
Additionally, we use the {\fontfamily{qcr}\selectfont sckit-learn} package \cite{scikit-learn} to optimize the hyperparameters as in the implementation of Algorithm 2.1 in \cite{gpr06}. 

This method is non-parametric because no direct model ansatz is used.
Note however that a choice for the covariance function needs to be made.\\
%Although we can set the parameters of \GPR and its mathematical background follows some logic, 
%the optimization of the hyperparameters involves a machine learning process.
%Additionally, this method is known as nonparametric because the computational time grows with more training points.
%Thus, in other studies, neural networks have been employed to optimize the hyperparameters using the integration of the hybrid Monte Carlo \cite{gpr96}. \\

\textbf{The covariance functions}

%Here we give examples of covariance functions.
In statistics, covariance expresses how likely two random variables change together \cite{guide_book}.
Various choices of covariance functions which are usually called kernels $k (\vec{x}, \vec{x}')$ are discussed in more detail in Ref.~\cite{scikit-learn} and \cite{gpr06}.
In this study, we compare the two most commonly used kernel functions in \GPR: the squared exponential kernel and the Mat\'{e}rn kernel explained below.
\begin{enumerate}
%Depending on the approximated function or the data set, the user can choose an appropriate kernel.
	\item The squared exponential kernel (SE) is a standard kernel for Gaussian processes: 
	\begin{equation}
	\label{SE}
	k_{SE}(r) = \exp \bigg(\frac{-r^2}{\ell^2}\bigg),	
	\end{equation}	
	with $r$ defined in Eq.~\ref{eq:gpr_r} and $\ell$ is the length-scale.
	\item The Mat\'{e}rn class of kernels is named after a Swedish statistician, Bertil Mat\'{e}rn and has less smoothness than the SE kernel. 
	The Mat\'{e}rn kernel is given by:
	\begin{equation}
	\label{Matern1}
	k_M(r) = \frac{2^{1-\nu}}{\Gamma(\nu)} \bigg(\frac{\sqrt{2\nu}r}{\ell}\bigg)^\nu K_\nu \bigg(\frac{\sqrt{2\nu}r}{\ell}\bigg),
	\end{equation}
	where $K_\nu$ is a modified Bessel function \cite{Abramowitz65}, $\Gamma$ is the gamma function and $\nu$ is usually half-integer. 
	Common choices of $\nu$ are $k_{\nu=3/2}$ and $k_{\nu=5/2}$.
	\be	
	\label{Matern32}
	k_{\nu=3/2} (r) = \bigg(1 + \frac{\sqrt{3}r}{\ell}\bigg) \exp \bigg(-\frac{\sqrt{3}r}{\ell}\bigg),
	\ee
	\be	
	\label{Matern52}
	k_{\nu=5/2} (r) = \bigg(1 + \frac{\sqrt{5}r}{\ell} + \frac{5r^2}{3 \ell^2}\bigg) \exp \bigg(-\frac{\sqrt{5}r}{\ell}\bigg).
	\ee	
	The Mat\'{e}rn kernel is a generalization of the radial basis function kernel. For $\nu=1/2$, it
	reduces to exponential kernel and $\nu=\infty$ reduces to the SE kernel.
	We use the Mat\'{e}rn kernel with $\nu=3/2$ in our analysis.
\end{enumerate}

\item \textbf{Artificial neural networks}
\label{sec_ann}

Artificial neural networks (\ANNs) as computing systems are inspired by emulating the work of brains to learn complex things and to find patterns in biology.
In machine learning algorithms, \ANN has been widely used as a framework to perform advanced tasks 
such as pattern recognition \cite{imageprocessing_ann}, forecasting \cite{forecast_ann}, and many other applications in various disciplines \cite{appls_ann}.
This framework works analogously to brains: it receives some inputs, processes them, and yields some output \cite{dlgoodfellow}.
%\ANN has become competitive to conventional regression and statistical models.

In this study, we employ \acp{ANN} or feedforward networks as the simplest neural networks architecture to perform interpolation. 
The feedforward network with hidden layers can approximate of any function which is known as the universal approximation theorem \cite{dlgoodfellow, Hornik}.
This class is called \emph{feedforward} because the information flow from the input to the output and the connection between them does not form a cycle (loop).
In our case, the inputs are the waveform's parameters $\vec{\lambda}$ and the output is the predicted value of $A(t_i;\vec{\lambda})$ or $\phi(t_i;\vec{\lambda})$.
We define \emph{hidden layer} as a layer between the input and the output of \ANN \footnote{In some references, the input layer is counted as the first hidden layer. Here we use the definition of \emph{hidden layer} as a layer between the input and the output layer}.

Four types of commonly used \acp{ANN} are:
\begin{itemize}
\item Single-layer perceptron \\
In a single-layer perceptron, the inputs are weighted and fed directly to the output. Hence, the single-layer perceptron is the simplest neural network system.
\item Multi-layer perceptron  \\
In \MLP, there is at least one hidden layer between the input and the output layer, 
where each neuron in each layer is connected to another neuron in the following layer. 
%\MLP takes the value of each parameter as its input without further process.
\item Radial basis function network \\
This class has the same workflow and architecture as the MLP with input, hidden layers and output,
where each neuron is connected directly to the following layer.
The only difference is the input, where the \ac{RBFN} uses the Euclidean distances with respect to some origin as its input and Gaussian activation functions \cite{orr96}. 
\item Convolutional neural network \\
The feedforward convolutional neural network is commonly used to train neural network for visual analysis.
Convolutional neural networks use convolution in place of general matrix multiplication in at least one of its layers \cite{dlgoodfellow}.
%These networks play an important role in practical applications.
\end{itemize}

We employ \MLP as one of the simplest architectures to perform function approximation \cite{Hornik, Sonoda}. 
Fig.~\ref{fig:diagram_ann} shows the illustration of the network architecture used in this study.

\begin{figure}
\begin{center}
\includegraphics[width=0.5\textwidth]{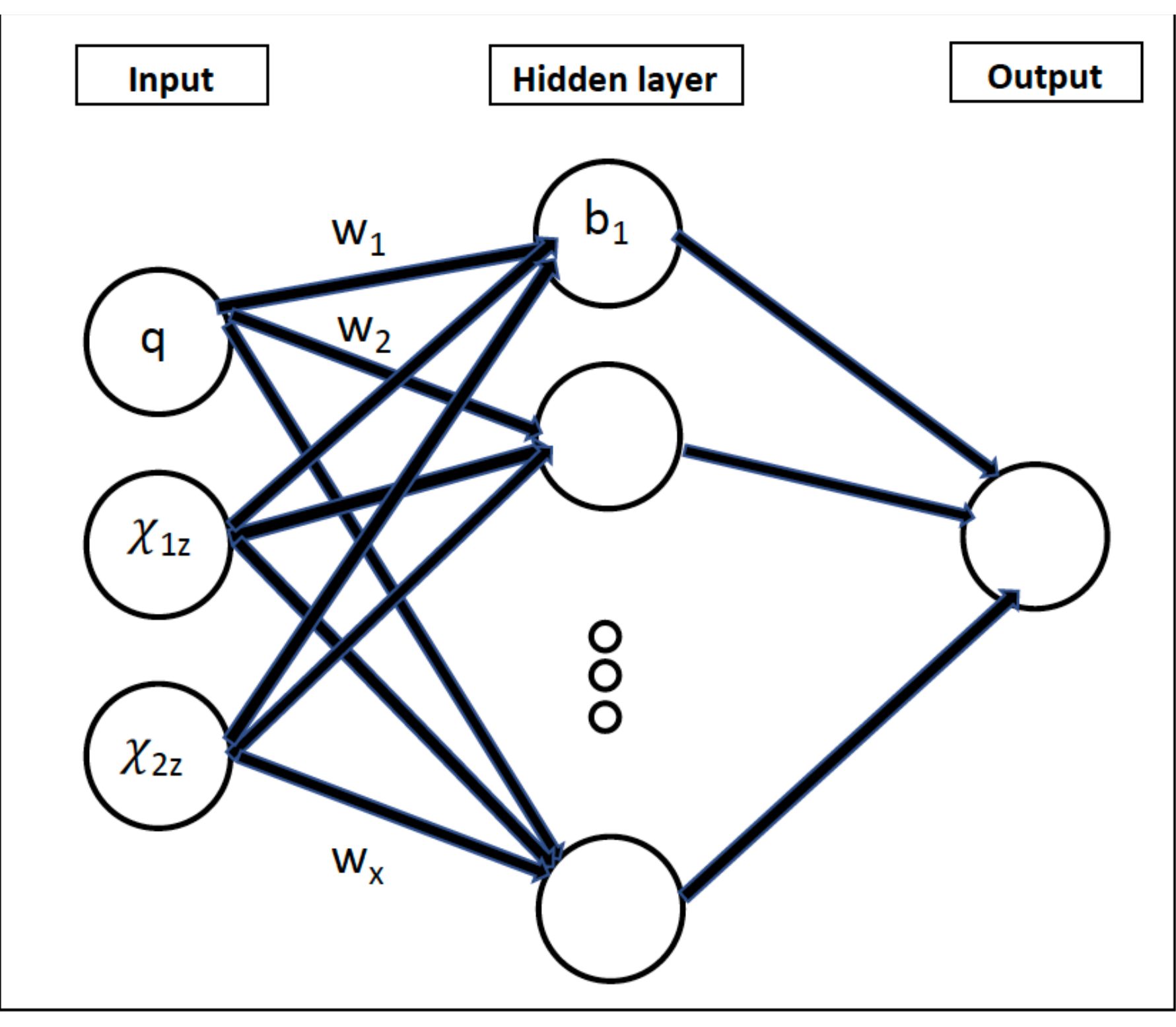}
\caption{Diagram of \ANN architecture used for three-dimensional interpolation in this study. 
The circles represent the neurons and we indicate weigths $w_i$ along neuron connections and biases $b_i$.%, whereas the connection between layers are called the weights ($w$) and bias $b$. 
We employ two layers in the hidden layer part of the diagram.
The same architecture is used for the seven-dimensional case, where the input contains seven neurons that depend on the seven parameters.}
\label{fig:diagram_ann}
\end{center}
\end{figure}

In Fig.~\ref{fig:diagram_ann}, each layer consists of a finite number of neurons. 
Each neuron in each layer is connected to the subsequent layer and the previous layer which are generally called \emph{links} or \emph{synapses}.
The workflow of \MLP is explained as follows:
\begin{enumerate}
\item Define the input as $x_{ij}$, where $i$ is the index of the layers. 
Starting at $i=0$ at the input layer, and $j$ indexes the neurons in a layer.
Thus, with $x_{0j}$, $j=1,2,3$ corresponds to $q, \chi_{1z}, \chi_{2z}$ respectively.
\item The $k$-th neuron of the $({i+1})$-th layer receives the value of $x_{ij}$ from the ${i}$-th layer multiplied by the weight $w_{ijk}$.
These products are then summed over all links from the $i$-th to the $(i+1)$-th layer.
\item A bias or shift $b_{ik}$ is added to the above value and an activation function $\sigma$ is applied to the final result.
In this study, we use the Rectified Linear Unit (ReLU) \cite{ReLU11} because it faster than other functions such as \emph{sigmoid} and \emph{tanh} and it is commonly used in other studies.
ReLU is mathematically expressed by the following equation:
\be
\sigma(z)=\textrm{max}(z,0),
\ee
\end{enumerate}
and the \MLP procedure is expressed by the following relation:
\be
\label{eq:ann}
x_{i+1,k}=\sigma \left(\sum_j w_{ijk}x_{ij}+b_{ik}\right)
\ee

We vary the number of neurons in the first hidden layer between 2 to 2000 for the three-dimensional data sets 
and 2 to 5000 for the seven-dimensional data sets.
We then set the number of neurons in the second hidden layer identical to the first hidden layer.
For each network and training data set, we compute 
mean squared error and the mean absolute error (see \cite{chollet2015keras}) of $A(t)$ and $\phi(t)$, respectively.

To train the networks, the training data is separated into several \emph{batches},
where each batch contains the same number of data samples.
Each batch is then passed through the networks (see Eq.~\ref{eq:ann}). 
When each data sample in the training set has had an opportunity to pass the networks a single time,
this is known as an \emph{epoch}.
The number of epochs affects the learning of the networks, i.e., the higher the epoch, the better the learning.
In this study, we set our batch size to five and train them through one thousand epochs.

The networks compute the \emph{loss functions} during each epoch.
The \emph{loss functions} measure the errors or inconsistency between the predicted value and the true data.
In this study, we employ the \emph{mean squared error} loss function for $A(t)$ and the \emph{absolute error} for $\phi(t)$ respectively (see Ref.~\cite{chollet2015keras}).

Training neural networks means that we minimize the loss functions so that our predicted values are as close as possible to the true values \cite{Nunes17}.
To minimize the loss functions, the networks adjust learnable parameters, i.e., the values of the weights and biases of the model.
In most cases, the minimization cannot be solved analytically, but can be approached with \emph{optimization algorithms}.

During optimization, the network learns the values of weights and biases of the previous epoch and calculates its loss functions.
Subsequently, it adjusts the values of weights and biases in the next epoch so that the loss functions become smaller.
One way to minimize the loss functions is to compute the gradient values with respect to the learnable parameters.
In this study, we employ \emph{Adam} \cite{Adam17} as the optimization algorithm.
Adam is a popular algorithm in deep learning due its robustness (see Ref.~\cite{Adam17} for more detail).
%To train the networks, the $N$ training data is separated into several groups or \emph{batches}, 
%where each batch contains the same number of data samples.
%Each batch is then passed through eq.~\ref{eq:ann} until all the batches have been processed
%in a single time (known as \emph{epoch}).
%Evaluation is ran at the end of every epoch to compute the data fit.
%The larger the value of epoch, the better the approximation is expected.
%The computational time to train this networks is recorded as the training time.

Following the above procedure, a model is then saved at the end of the run and evaluated through the test data.
We then compute the accuracy and execution time of this process similar to other methods.
We employ {\fontfamily{pcr}\selectfont Keras} \cite{chollet2015keras} and {\fontfamily{pcr}\selectfont TensorFlow} \cite{tensorflow} to perform this computation.

\end{enumerate}

%===============================
%		RESULT
%===============================

\section{Results}
\label{sec:result}
%\todo{
%\begin{itemize}
%\item what makes one method is faster than the other/more accurate?
%\end{itemize}
%Summarize the advantages and disadvantages of each method.
%How one should prepare the data before performing interpolation
%}

In this section, we show results for accuracy and computational time for different regression methods.
We apply methods to the three-dimensional and seven-dimensional data sets defined in sec.~\ref{sec:interpmethods}.

\subsection{Three-dimensional case}
\label{sec:threedimensions}

We investigated the results for aligned spin waveforms with 
parameters $q, \chi_{1z}$, and $\chi_{2z}$.
Training points were given on a regular grid. 
We placed the same number of points equally spaced to each other for each parameter (see sec.~\ref{sec:preparedata}).  
Hence the total number of training points is proportional to the number of training points per dimension cubed.
We then varied the number of training points in each dimension from five to eleven which corresponds
to a total number of training points of 125 to 1331.
%Hence, our total number of training points lies within the range of the training points employed to build the NRSurrogate model.
We distributed 2500 test points randomly (see section \ref{sec:preparedata}).
These test points are located inside the same domain covered by the training points.
Hence, we do not test how well the methods perform for extrapolation.

We calculated relative errors (in percent) for the amplitude $A(t)$:
\begin{equation}
\label{relative_error}
\varepsilon_{re}=\frac{\sum_i^N |A_{\mathrm{pred}}^i (t)-A_{\mathrm{true}}^i (t)|}{\sum_i^N |A_{\mathrm{true}}^i(t)|} \times 100.
\end{equation}
%Since the phase error has physical intuition in unit radian,
%we compared the errors as the average of the absolute error with the following formula: 
The phase error is an important diagnostic to measure the accuracy of \GW waveform models.
Therefore, we consider the absolute phase error (in radians)
\begin{equation}
\label{abs_error}
\varepsilon_{ae}=\frac{1}{N}\sum_i^N |\phi_{\mathrm{pred}}^i (t)-\phi_{\mathrm{true}}^i (t)|.
\end{equation}
$\varepsilon_{re}$ and $\varepsilon_{ae}$ are the relative error and the average of the absolute error, respectively, 
$A_{\mathrm{pred}} (t)$ and $\phi_{\mathrm{pred}} (t)$ are the predicted results of the amplitude and phase regression respectively, 
and $A_{\mathrm{true}} (t)$ and $\phi_{\mathrm{true}} (t)$ are their true values.

\begin{figure*}[h]
\centering
     \includegraphics[width=0.44\linewidth]{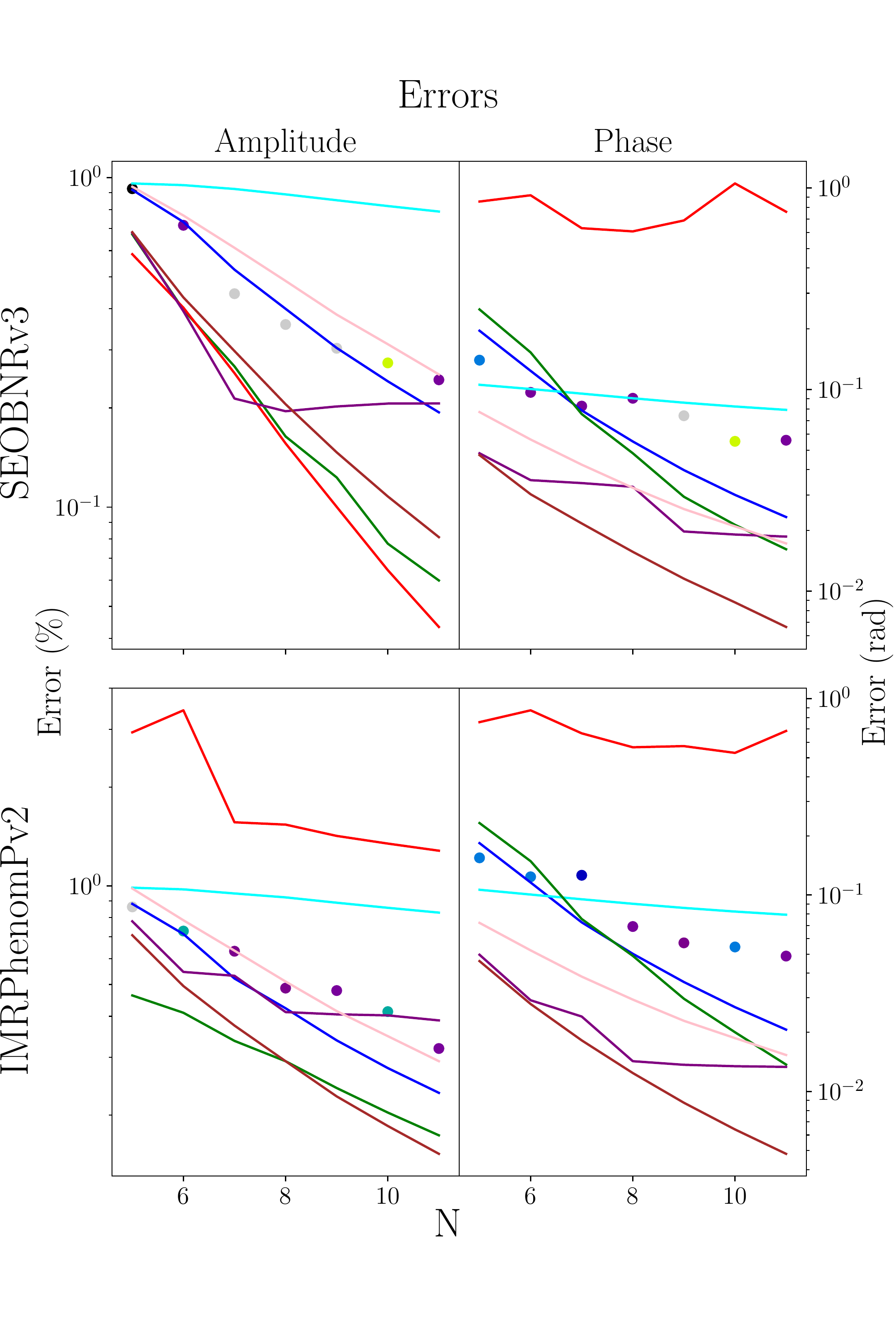}\hfil
    \includegraphics[width=0.28\linewidth]{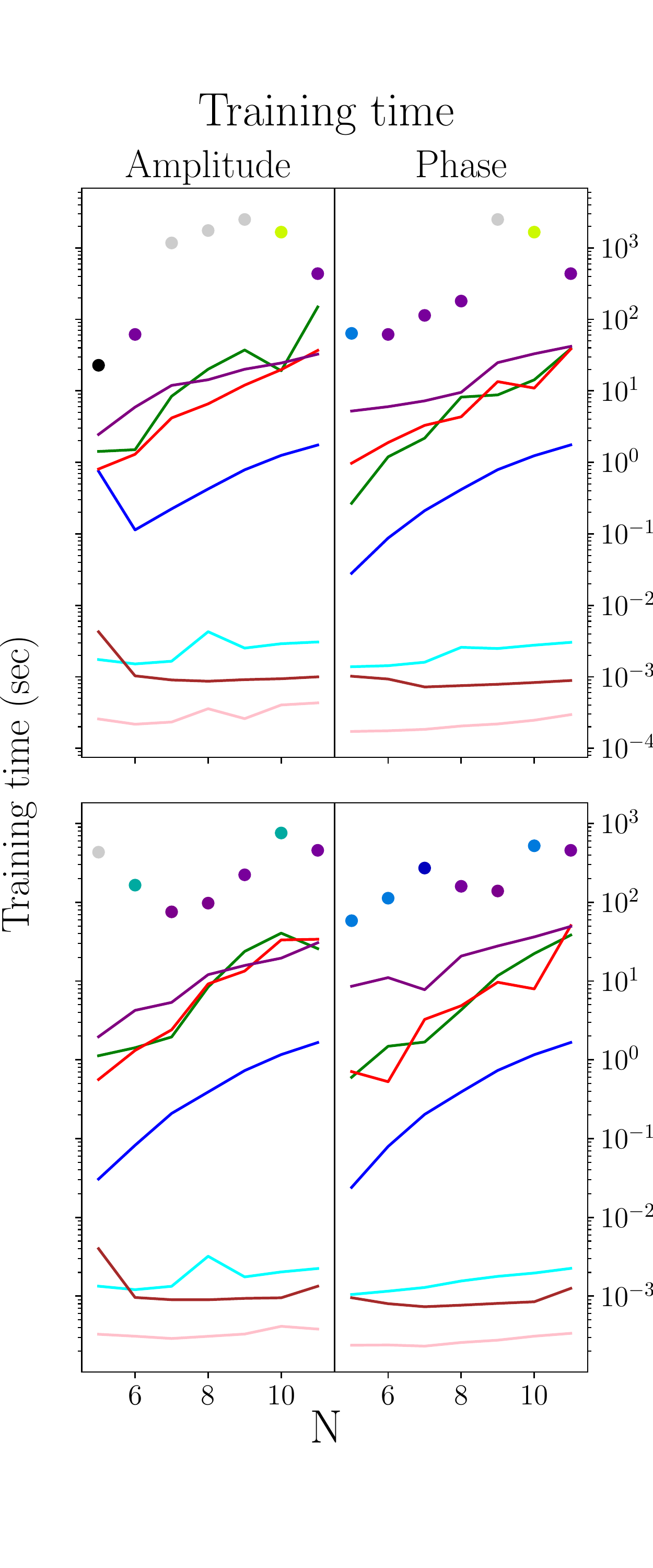}\hfil
    \includegraphics[width=0.28\linewidth]{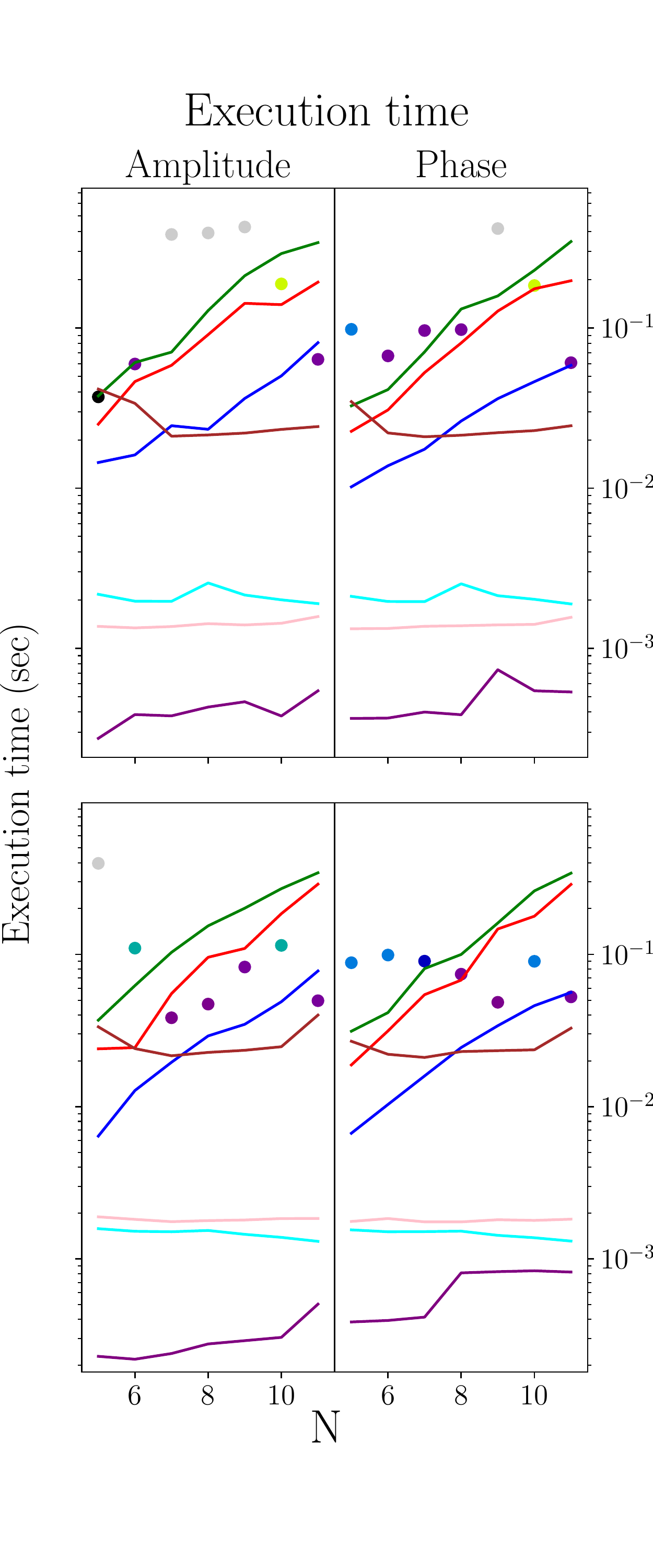}\par\medskip
     \includegraphics[width=0.9\linewidth]{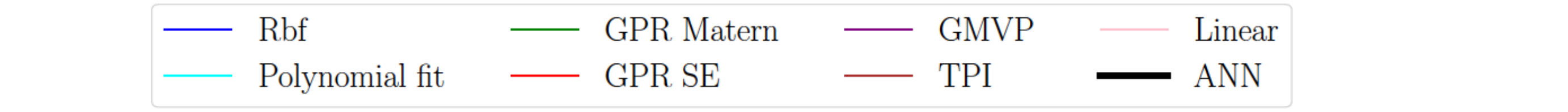}\par\medskip
    \includegraphics[width=0.5\linewidth]{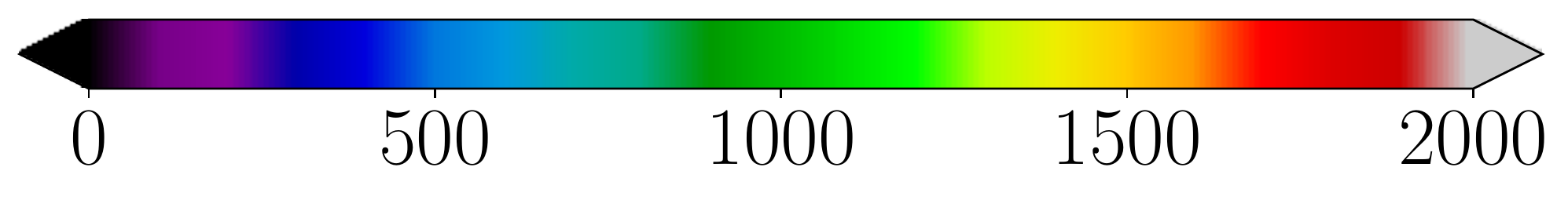} 
\caption{ The three-dimensional interpolation results at t=-3500M. Top:{\fontfamily{pcr}\selectfont SEOBNRv3}, 
bottom:{\fontfamily{pcr}\selectfont IMRPhenomPv2}. 
The $x$-axes show the number of training points in each dimension, $N$, and the $y-$axes 
show the errors, training, and indicated execution time as on the labels of the panels. 
Left: errors of the amplitude and phase respectively,
middle: training time in (seconds), and right: execution time (seconds).
Different colors represent different interpolation methods as shown in the shared legend.  
The colored circles show \ac{ANN} results, where different colors represent the number of neurons per layer in a double layer ANN 
as shown in the corresponding color bar.}
	\label{fig:3dplots}  
\end{figure*}

\begin{figure*}[h]
\centering
     \includegraphics[width=0.44\linewidth]{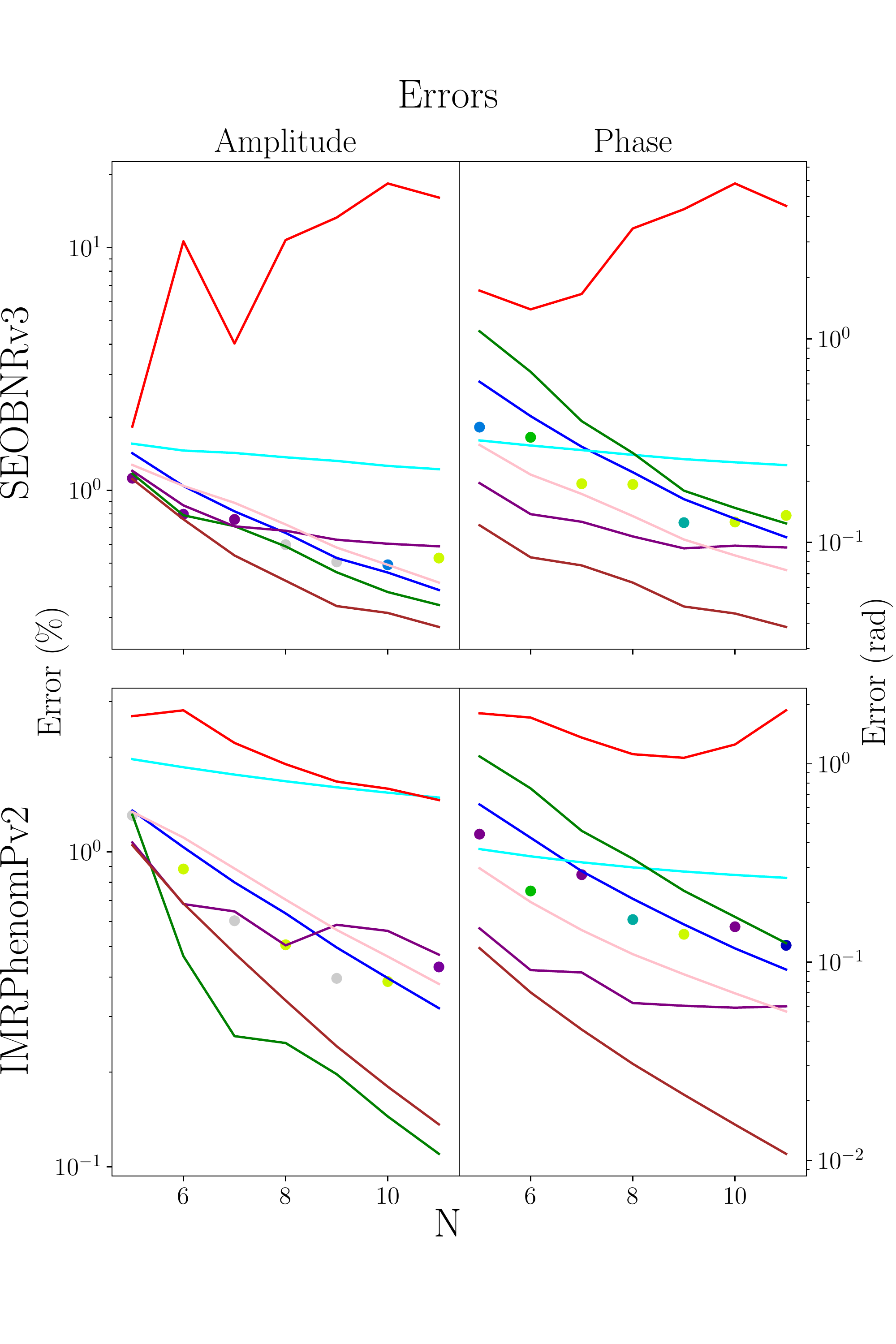}\hfil
    \includegraphics[width=0.28\linewidth]{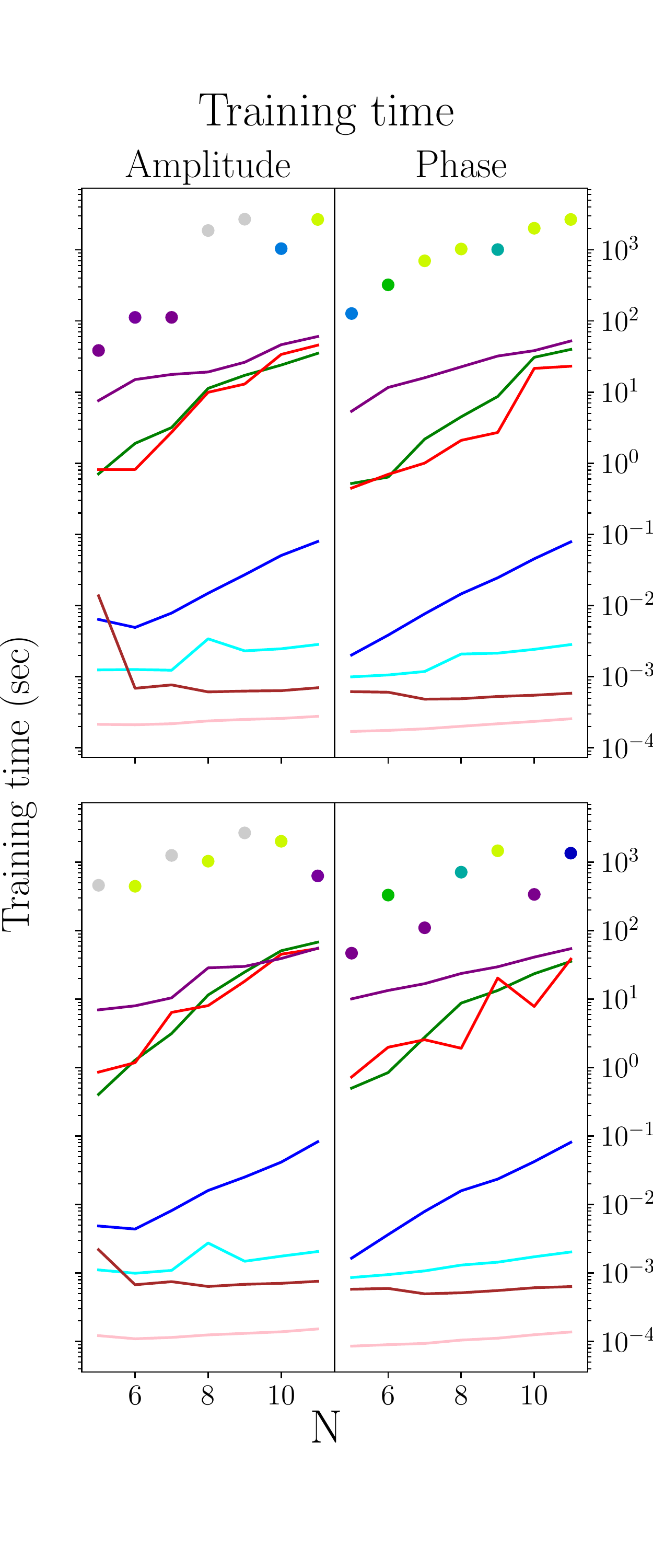}\hfil
    \includegraphics[width=0.28\linewidth]{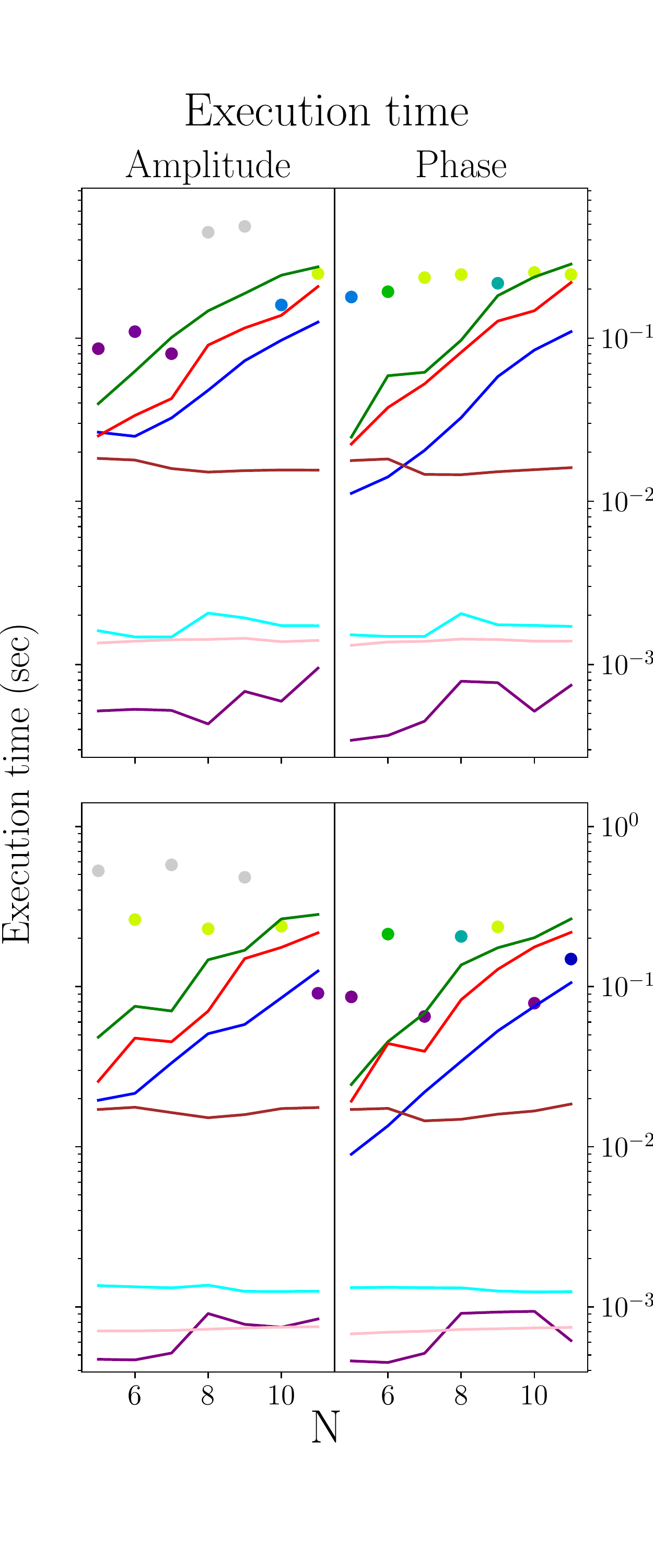}\par\medskip
     \includegraphics[width=0.9\linewidth]{legend_3d.pdf}\par\medskip
    \includegraphics[width=0.5\linewidth]{3d_colorbar.pdf} 
\caption{ The three-dimensional interpolation results at t=-50M. Top:{\fontfamily{pcr}\selectfont SEOBNRv3}, 
bottom:{\fontfamily{pcr}\selectfont IMRPhenomPv2}. 
The $x$-axes show the number of training points in each dimension, $N$, and the $y-$axes 
show the errors, training, and indicated execution time as on the labels of the panels. 
Left: errors of the amplitude and phase respectively,
middle: training time in (seconds), and right: execution time (seconds).
Different colors represent different interpolation methods as shown in the shared legend.  
The colored circles show \ac{ANN} results, where different colors represent the number of neurons per layer in a double layer ANN 
as shown in the corresponding color bar.}
	\label{fig:t503dplots}  
\end{figure*}

Subsequently, we investigated the computational time taken to evaluate each interpolation method.
Here we define the \emph{training time} as the time to compute the interpolant and the 
\emph{execution time} being the time to compute the 2500 interpolation points following our test set.
Furthermore, we define \emph{total time} as the sum between the training time and the execution time, i.e., the entire process to perform interpolation for 2500 points.
The comparison results in the early inspiral ($t=-3500M$) are shown in Fig.~\ref{fig:3dplots}, 
whereas the results at $t=-50M$ are shown in Fig.~\ref{fig:t503dplots}.
We now discuss the results shown the results for different regression methods.

\begin{enumerate}

\item \textbf{Traditional interpolation and fitting methods \& \GPR}

We expect that the key quantities for two waveform models, {\fontfamily{pcr}\selectfont SEOBNRv3} and {\fontfamily{pcr}\selectfont IMRPhenomPv2} 
agree quite well in the early inspiral. 
The error in $A(t)$ and $\phi(t)$, decreases with more training points for both models. 
This result is expected as we populate our parameter space with more points located on a regular grid.

For both quantities, we find that errors for different methods are similar between waveform models.
\GPR errors show a dependence on the kernel choice.
We first consider the amplitude errors. For {\fontfamily{pcr}\selectfont SEOBNRv3} the errors fall off in a similar way for either choice of kernel, whereas for {\fontfamily{pcr}\selectfont IMRPhenomPv2}
the error is much higher for the SE kernel compared to the Mat\'{e}rn kernel.
This is likely due to the higher level of noise in the {\fontfamily{pcr}\selectfont IMRPhenomPv2} data due to the inverse Fourier transformation. 

The SE kernel assumes a higher degree of smoothness in the data than the Mat\'{e}rn kernel. 
Similarly, we find for either waveform model that the SE kernel shows a higher phase error than the Mat\'{e}rn kernel.
	
\item \textbf{Artificial neural networks}

We now discuss errors for \acp{ANN} as indicated by the filled circles in Fig.~\ref{fig:3dplots}.
Here we compare the results of the double layer \ac{MLP} with various numbers of neurons.
By design, the double layer \ac{MLP} consists of one input layer, two hidden layers, and one output layer.
We set the number of inputs as the dimensionality of the parameter space and only produce a single output.
In the aligned spin case, our inputs are the parameters $q, \chi_{1z}$, and $\chi_{2z}$ and output is either $A(t)$ or $\phi(t)$. 
For the hidden layers, we varied the number of neurons between 2 and 2000 in the first hidden layer,
and set an equal number of neurons for the second hidden layer.

Thus, we obtained a set of errors as we modified the number of neurons in the \emph{hidden layers} 
for a fixed number of training points $N$ per dimension.
In Fig.~\ref{fig:3dplots}, we only show the results of the smallest errors for each training set.
In this plot, different colors of the circles correspond to different numbers of neurons as indicated by the color bar.
We note that the \ANN with the smallest error may not be the fastest one.
%One might expect that the \ANN with the largest value of neurons per layer us the most accurate one.

Regarding the computational time, the training time obviously grows with the number of neurons per layer. 
However, we argue that there is no guarantee that many neurons yield smaller error than fewer neurons.
In fact, too many neurons lead to overfitting and too few neurons lead to underfitting.
We could reduce overfitting by activating the \emph{Dropout} function in {\fontfamily{pcr}\selectfont Keras},
 \emph{Dropout} removes the result from a selected number of neurons randomly.
However, we prefer to not include an additional stochastic element and do not include \emph{Dropout} in this study.

Next, we compare execution times.
Execution time is relatively similar between
the \GPR, \RBF, \TPI, and \ANN methods.
Other traditional methods such as linear, polynomial fit and \GMVP, and linear interpolation 
are faster.

To ensure a fair comparison between all methods, we explored the performance on the same machines 
(2x Intel Xeon E5-2698 v4) with 20 CPU cores, 256 Gigabytes of RAM, and 1x HDD (1TB, 6Gbps) of storage.

Due to the limited scope of our study, we only investigate results for the double layer \ANN.
This leaves tuning parameters and architectures to be explored in future studies.
A possible way to reduce training and execution times is to use on GPUs instead of CPUs. 

\end{enumerate}
%Although the single layer \ANN can perform interpolation, the weight is computed linearly.
%Thus, it is mathematically rather similar than other interpolation methods.
%Besides, more neurons in single layer perform comparable to less neurons in a double layer.

%We then focused on double layer \ANN of \ac{MLP} and varied the number of neurons.
%In this case, we use the same number of neurons in each layer from 2 to 2000.
%In Fig.~\ref{fig:3dplots}, we show the result of the number of neurons with the smallest errors in each training set
%as shown in its colorbar.

%The result of the three dimensional interpolation is shown in Fig.~ \ref{fig:3dplots}.
%Our finding is that more neurons need a longer training time, but there is no guarantee that the result performs better than
%less neurons.
%The execution time is comparable as shown in the same plot.
%However, a 

Finally, we discuss results for training times.
The training time for \RBF and \GPR rise proportionally with the number of training points.
In \RBF, this is caused by the least-squares-fit computation that takes a longer time with more training points.
For \GPR, the training time goes as $\mathcal{O}(N^3)$ with $N$ the number of training points 
as explained in Sect.~\ref{sec:interpmethods}.
Polynomial fit, \TPI and linear interpolation do not depend strongly on the size of the training set and their training time is relatively fast.

For both models, \ANN yields comparable errors and execution times as other interpolation methods, but
 generally with longer training time than other methods.
Several methods have execution times that are independent of the size of the training set for a fixed order of approximations.
This includes \TPI, linear interpolation, polynomial fit, and \acp{ANN}.

Combining all the results at $t=-3500M$ and at $t=-50M$, we found that the errors are generally larger in noisy data.
We also found that the methods with longer training time do not always yield a better result than the methods with less training time (see Fig.~\ref{fig:t503dplots}).

Using too many neurons in the hidden layers may cause problems such as overfitting.
It occurs when the networks have too much capacity to process information such that the amount of information in the training set is not enough
to train the networks \cite{HeatonNN}. Hence, the number of neurons must be set such that there are not too few or not too many.
The selection however, depend on the architecture of the networks and the \emph{hyperparameters}.

\subsection{Seven-dimensional case}
\label{sec:sevendimensions}

In seven dimensions, we distribute the training points randomly in each dimension.
The main reason for this placement is to avoid the \emph{curse of dimensionality} as explained in the previous section.
Similarly to the three-dimensional case, we investigate training sets of different sizes, from 500 to 3000 points.
As discussed in Sec.~\ref{sec:preparedata}, the seven-dimensional case has a narrower range of mass ratio ($1\leq q \leq 2$) than the three-dimensional one ($1\leq q \leq 10$)
and full-spin range.

We construct a single test set with 2500 points distributed randomly and located within the parameter ranges.
Some of the test points may be outside the domain covered by the training points.
This means that our results may contain a small extrapolation.

Since \TPI and linear interpolation require regular grid training points, we do not include them in our analysis.
For other methods, we employed the same settings (kernels, hyperparameters, degree) as in the three-dimensional case.

\begin{figure*}[h]
\centering
    \includegraphics[width=0.44\linewidth]{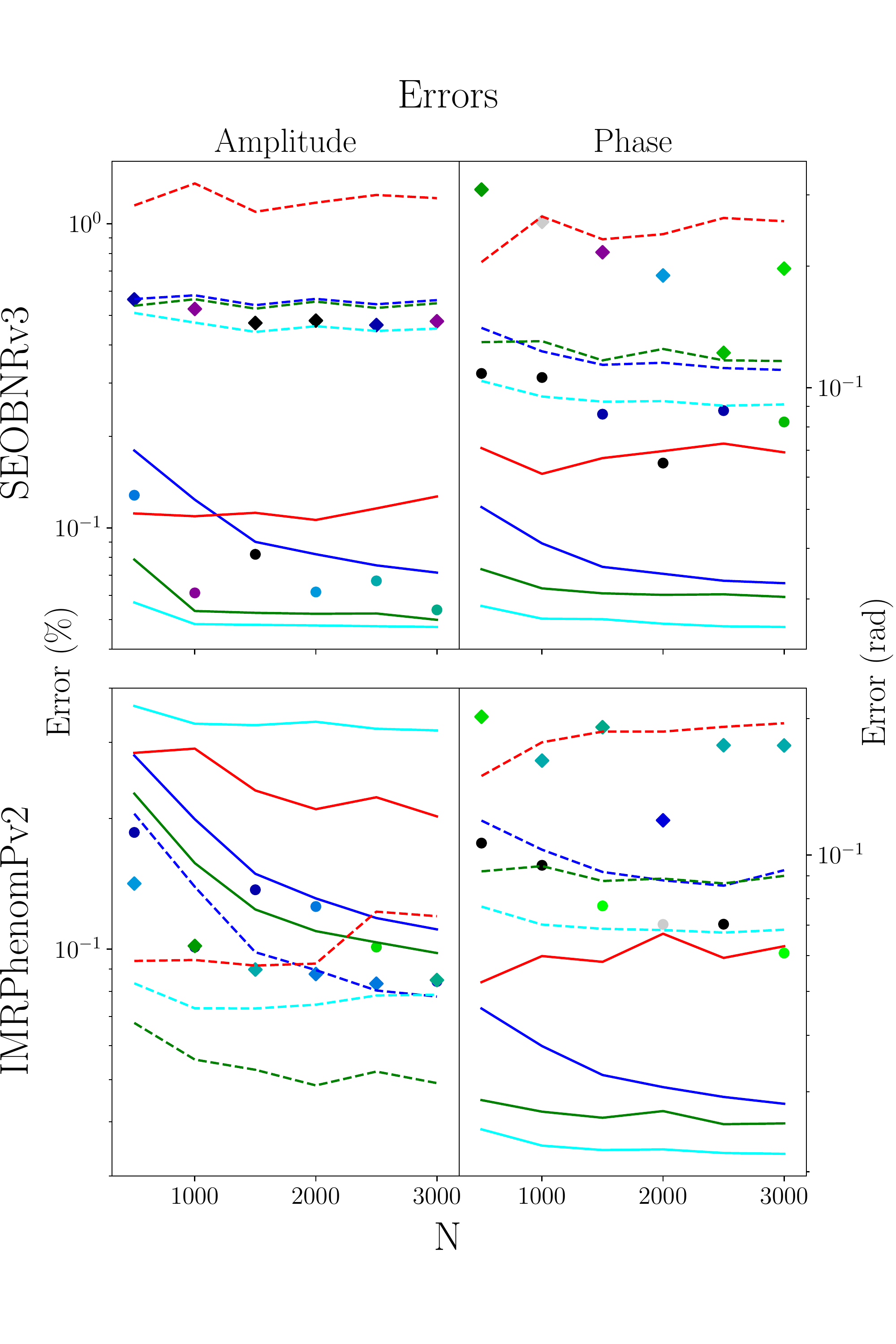}\hfil
    \includegraphics[width=0.28\linewidth]{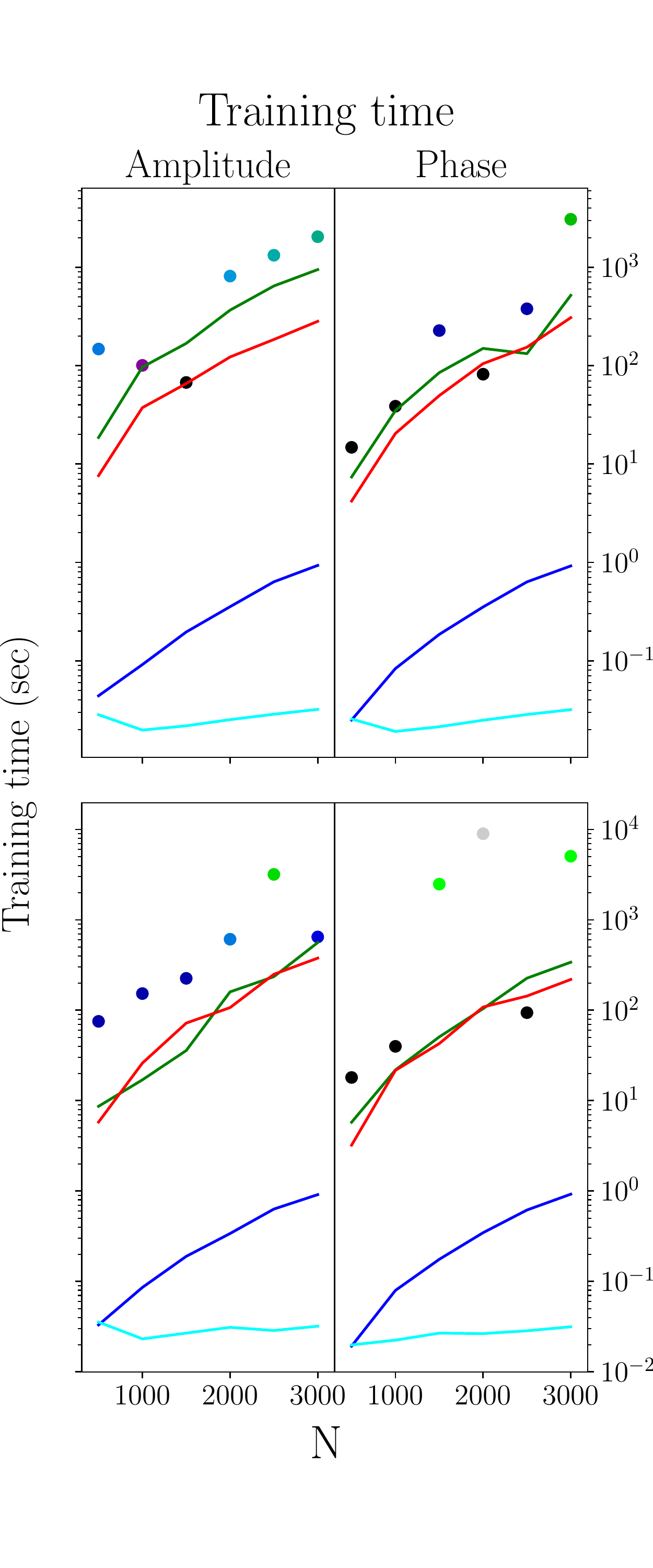}\hfil
    \includegraphics[width=0.28\linewidth]{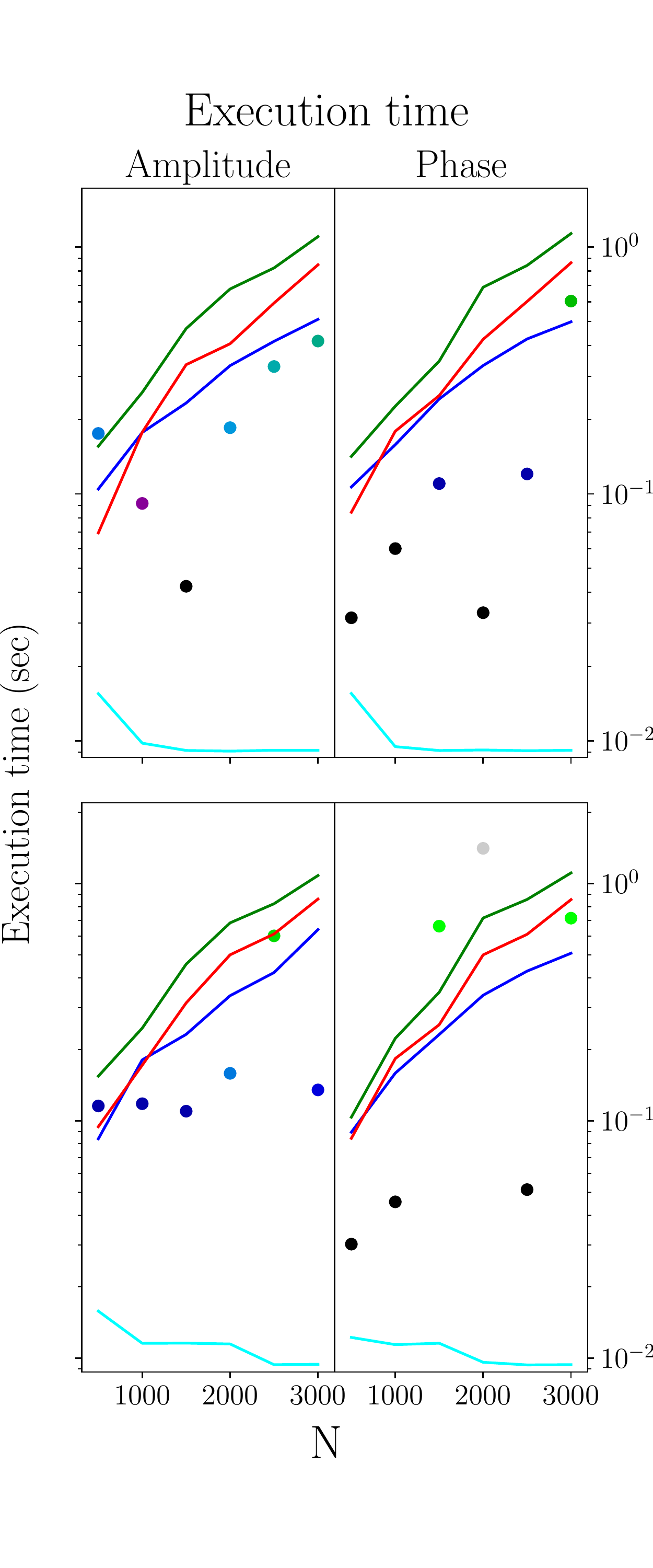} \par\medskip
     \includegraphics[width=0.9\linewidth]{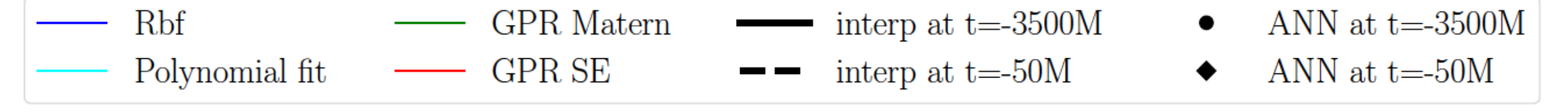} \par\medskip   
    \includegraphics[width=0.5\linewidth]{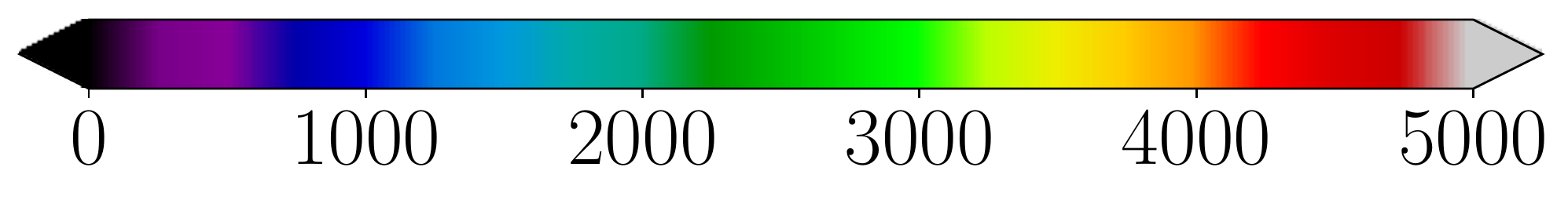}        
\caption{The seven-dimensional interpolation results. Top:{\fontfamily{pcr}\selectfont SEOBNRv3}, 
bottom:{\fontfamily{pcr}\selectfont IMRPhenomPv2}. 
The $x$-axes show the number of training points $N$ and the $y$-axes shows the errors, training, and execution time as shown on the plot. 
Left: errors of the amplitude ($A(t)$) and phase ($\phi(t)$) respectively,
middle: training time in unit seconds, and right: execution time in unit seconds.
The solid lines show the results at $t=-3500M$ and the dashed lines for $t=-50M$.
Different colors represent different interpolation methods as shown in the shared legend.  
The colored circles correspond to the results of ANN at $t=-3500M$ and the colored diamonds for $t=-50M$. 
Different colors represent a different number of neurons on a  double layer ANN 
as shown in the corresponding color bar.}
	\label{fig:sevendplots}
\end{figure*}

We built the architecture of \ANN in a similar way as before.
The results of the seven-dimensional case for different interpolation methods ($t=-3500M$ and $t=-50M$) are shown in Fig.~\ref{fig:sevendplots}.

We observed that errors of {\fontfamily{pcr}\selectfont SEOBNRv3} are not significantly different than the corresponding three-dimensional results.
Furthermore, the errors of this model at $t=-50M$ are higher than at $t=-3500M$ in a similar way as in three dimensions.
 
Surprisingly, the relative amplitude errors for {\fontfamily{pcr}\selectfont IMRPhenomPv2} (top left plots) 
in the late inspiral are smaller than in the early inspiral in contrast to {\fontfamily{pcr}\selectfont SEOBNRv3}.
%It is caused by different functionalites between points in {\fontfamily{pcr}\selectfont IMRPhenomPv2} compare to {\fontfamily{pcr}\selectfont SEOBNRv3} at $t=-50M$.
The $A(t)$ quantity of {\fontfamily{pcr}\selectfont IMRPhenomPv2} is smoother at $t=-50M$ than at $t=-3500M$.
We emphasize that both models, {\fontfamily{pcr}\selectfont SEOBNRv3} and {\fontfamily{pcr}\selectfont IMRPhenomPv2} have comparable amplitude values at $t=-50M$ and at $t=-3500M$.

In the early inspiral ($t=-3500M$), both waveforms agree well, similar to the three-dimensional case.
Hence, the percent errors are not significantly different as shown in the same plot.

The phase errors were computed as absolute errors (see Eq.~\ref{abs_error}).
We find that the phase errors for {\fontfamily{pcr}\selectfont SEOBNRv3} and {\fontfamily{pcr}\selectfont IMRPhenomPv2} are comparable.
Furthermore, the late inspiral errors are higher than the early inspiral as the data fluctuates more.
In Fig.~\ref{fig:sevendplots}, we observe a similar behavior for the training time as in three dimensions, 
where higher training time was found for \GPR, \ANN, and \RBF.
This is caused by the same factors as explained in the three-dimensional case.
For the execution time (right panel), we found that the more complex methods take longer time than the simpler methods.
For \RBF and \GPR this is due to their dependence on the size of the training set.
Interestingly, the execution time for \acp{ANN} is faster than \GPR and \RBF.
This is because \ANN picks the optimum weights and biases during the training and its execution time does not depend on the number of training points in the data.

We remind the reader that we set the parameter space of the seven-dimensions analysis narrower in mass ratio than the three-dimensions.
Hence, the errors should not be compared directly to the three-dimensional case.
For the same parameter ranges, the seven dimensional case yields errors up to 100 times larger for the $A(t)$ and 15 times larger for the $\phi(t)$.
The order of accuracy does not significantly change, where the best accuracy in this range is obtained by polynomial interpolation.

Overall, we found that in some cases, a simple method such as polynomial fit yields lower errors
and performs faster than the more complex methods.

%===============================
%		DISCUSSION
%===============================

\section{Discussion and conclusion}
\label{sec:discussion}

Various approximation methods play important roles in building gravitational waveform models.
Methods with high accuracy, low complexity, and fast computational time are needed for current and future applications.
In this paper, we presented a comparative study of interpolation, fitting and regression methods applied to precessing and aligned \BBH systems.
Precessing \BBH model depends on seven key intrinsic parameters ($q, \vec{\chi}_1, \vec{\chi}_2$), whereas the aligned model depends on three parameters ($q, \chi_{1z}, \chi_{2z}$).

We generated the data sets in the time domain using two waveform models: {\fontfamily{pcr}\selectfont SEOBNRv3} (originally built in the time domain)
and the inverse Fourier transform of {\fontfamily{pcr}\selectfont IMRPhenomPv2} (originally built in frequency domain).
The full waveforms were transformed into a precession adapted frame where we extracted two quantities: amplitude $A(t)$ and phase $\phi (t)$ as explained in section~\ref{sec:preparedata} to perform a comparative study.
For each key quantity, we picked two points in time, $t=-3500M$ in the inspiral for the smoother data set and $t=-50M$ near merger for the more irregular data.
We employed this procedure on different numbers of training sets and used different approximation methods.

We split approximation methods into two categories: traditional methods and machine learning mehods (see Sect.~\ref{sec:interpmethods}).
The traditional methods consist of linear interpolation, polynomial fits, radial basis function, \GMVP, and \TPI.
Since linear interpolation and \TPI package require a regular grid, we do not include them in the seven dimensional analysis.
Furthermore, we investigated machine learning methods such as \GPR and \ANN.
For \GPR, we compared two kernel functions: the square exponential kernel and the Mat\'{e}rn kernel. 
We took the mean results of each kernel and compared them against other methods.
For \ANN, we focused on networks with two hidden layers and varied the number of their neurons. 

We computed the relative errors for $A(t)$ and the absolute errors for $\phi(t)$.
To validate the result, we generated 2500 test points distributed randomly within the same parameter space.
The comparison results of different methods in accuracy, training time and execution time (in second) are presented in Sec.\ref{sec:result}.

We found that all methods perform better with more training data.
Furthermore, we compared the performance of the same method in a set of smoother data and a set of more irregular data.
In general, we found that approximation methods perform better in smoother data as expected.
%Hence, we argue that data preparation are needed to smooth the data set before performing any approximation methods.
We recommend to use preprocessing methods to improve the smoothness of the data where possible which should increase the accuracy of regression results.
This preparation is crucial as any methods perform well with smoother data sets. 
Different accuracies are attained by different methods in handling the irregularities in the data.
We give a brief summary of different methods in Table.~\ref{tab:sum}.

\begin{table*}[b]
\footnotesize
\centering
   \begin{tabular}{llll}
\br
%\cmidrule(lr){2-4} \cmidrule(lr){5-7} \cmidrule(lr){7-7}
    Methods & Advantages & Disadvantages & Training \\
     &  &  &  time \\
\mr
  Linear (RGI)& standard {\fontfamily{qcr}\selectfont scipy} & needs regular grid & $\mathcal{O}(N$) \\[4pt]
  \TPI& robust and & needs regular grid & $\mathcal{O}(N^k)$ \\
   & high accuracy &  &  \\[4pt]
  \GMVP& irregular grid & complex &  \#basis function \\
   & fast execution time &  & \#error tolerance\\[4pt]
  Polynomial fit& irregular grid & Runge's phenomenon &  $\mathcal{O}(N)$ and \\
  & simple and fast & only univariate in {\fontfamily{qcr}\selectfont scipy} & \#polynomial degree\\[4pt]
  \RBF& {\fontfamily{qcr}\selectfont scipy}& high computational & $\mathcal{O}(N^3$) \\
   & irregular grid& complexity &  \\[4pt]
  \GPR& irregular grid& depends on the choice & $\mathcal{O}(N^3$) \\
  & can predict uncertainty & of kernel and hyperparameters & \\
  &  & complex &  \\[4pt]
  \ANN& irregular grid & complex & \#neurons \\
  & flexible architecture choices &  & \#hidden layers \\
\br
\end{tabular}
   \caption{Summary of features of the methods used in this study. We present the advantages, the disadvantages and the scaling complexity for each method.
   For linear interpolation, \TPI, \RBF, and \GPR the (training time) depends on the number of training points $N$ (and polynomial degree $k$).
   Other methods have different complexity scalings that affect their training time.}
   \label{tab:sum}
\end{table*}
%Methods with the best accuracy may require more computational complexity: longer training time and even execution time compare to other methods.
%The application of different methods depend on the goal of the study.
%If high accuracy is desired, users can decide how far the tolerance to the training time.
%This also applies if computational time is expected rather than accuracy.
%Computational complexity depend on the number of training sets.
%Furthermore, different methods have different features that useful in various studies.
For lower dimensions, simpler methods such as linear interpolation and \TPI provide good accuracy and speed.
However, these methods need a regular grid and therefore are less useful for high dimensional data sets as explained above.
For this situation, we found that polynomial fits are one of the simplest methods that offers a good combination between accuracy and speed.
Furthermore, polynomial fits have been used widely and can be coded manually making it reliable and easy.
The computational timing of polynomial fits depends on the number of parameters and the maximum polynomial degree.
Another method that can perform approximation of scattered data sets is \GMVP.
\GMVP which is based on polynomials can perform very well by setting error tolerance on its algorithm.
For lower dimensionality, \GMVP is computationally cheap.
However, as the number of parameters rise, the computational time to compute the interpolant with the same error tolerance grows significantly higher.
Therefore, we do not include this method in our analysis for the seven-dimensional case.

\RBF and \GPR are promising methods for scaterred data points.
\RBF has been integrated in a standard {\fontfamily{qcr}\selectfont scipy} package, making it easy for users.
\GPR computes the uncertainty of the predicted values. This feature is useful for future applications and cannot be found in other methods.
Furthermore, \GPR has been integrated in {\fontfamily{qcr}\selectfont sckit-learn} package \cite{scikit-learn}.
Both \RBF as \GPR have the freedom to choose suitable kernel functions and hyperparameters.
However, their speed depends on the number of training points cubed $\mathcal{O}(N^3)$.
Hence, these methods become inefficient for larger data set.

A simple \ANN can be used to perform regression for scattered data points.
Similar to \GPR, this method is more complex and depends on the choice of architecture and hyperparameters.
We showed that the the three-dimensional result of \ANN requires a longer training time with relatively comparable accuracy to other methods.
We argue that such complexity is less needed for lower dimensional parameter and users should use a more simpler methods that provide good accuracy and speed.
However, \ANN is highly versatile to solve problems in higher dimensions and is promising to be explored further.

One might expect that methods with higher complexity perform better than methods with lower complexity.
We find that this is not always the case.
A more complicated method does not guarantee that the results are always better or faster.
We find that simpler methods may yield smaller errors than more complex methods and perform faster in many cases.
Hence, we suggest that one should critically evaluate the performance of approximation methods and understand the features of the method that are necessary for the data of interest.
Simpler methods that perform better or at least equal to more complicated methods should be used as the first choice to avoid unecessary complexity.

%\todo{
%\begin{enumerate}
%\item How improvement can be made.
%\item Future studies.
%\item Suggestions to do the interpolation.
%\end{enumerate}
%}

\section*{Acknowledgments}
The authors would like to thank to Lionel London, Scott Field, Stephen Green, Chad Galley, Christopher Moore, Zoheyr Doctor, Rory Smith, Ed Fauchon-Jones, and Lars Nieder for useful discussions.
Computations were carried out
on the Holodeck cluster of the Max Planck Independent Research Group ''Binary 
Merger Observations and Numerical Relativity.`` This work was
supported by the Max Planck Society's Independent Research Group 
Grant. 

\section*{References}
\bibliography{paper}

\end{document}